\documentclass[preprints,article,accept,moreauthors]{mdpi}

\firstpage{1} 
\pubvolume{1}
\issuenum{1}
\articlenumber{0}
\pubyear{2024}
\copyrightyear{2024}
\datereceived{ } 
\daterevised{ } %
\dateaccepted{ } 
\datepublished{ } 
\hreflink{https://doi.org/} %

\usepackage{booktabs,makecell}

\Title{Partition Function Zeros of the Frustrated $J_1$-$J_2$ Ising Model on the Honeycomb Lattice}

\TitleCitation{Complex partition function zeros of the frustrated $J_1$-$J_2$ Ising model on the honeycomb lattice}

\Author{Denis Gessert $^{1,2}$\orcidA{}, Martin Weigel $^{3}$*\orcidB{} and Wolfhard Janke $^{1}$\orcidC{}}

\AuthorNames{Denis Gessert, Martin Weigel, Wolfhard Janke}

\AuthorCitation{Gessert, D.; Weigel, M.; Janke, W.}

\address{%
$^{1}$ \quad Institut für Theoretische Physik, Leipzig University, IPF 231101, 04081 Leipzig, Germany; denis.gessert@itp.uni-leipzig.de, wolfhard.janke@itp.uni-leipzig.de\\
$^{2}$ \quad Centre for Fluid and Complex Systems, Coventry University, Coventry CV1 5FB, United Kingdom; gessertd@uni.coventry.ac.uk\\
$^{3}$ \quad Institut für Physik, Technische Universität Chemnitz, 09107 Chemnitz, Germany; martin.weigel@physik.tu-chemnitz.de}

\corres{Correspondence: martin.weigel@physik.tu-chemnitz.de}

\abstract{We study the zeros of the partition function in the complex temperature plane (Fisher zeros) and in the complex external field plane (Lee-Yang zeros) of a frustrated Ising model with competing nearest-neighbor ($J_1 > 0$) and next-nearest-neighbor ($J_2 < 0$) interactions on the honeycomb lattice. We consider the finite-size scaling (FSS) of the leading Fisher and Lee-Yang zeros as determined from a cumulant method and compare it to a traditional scaling analysis based on the logarithmic derivative of the magnetization $\partial \ln \langle |M| \rangle /\partial\beta$ and the magnetic susceptibility $\chi$. While for this model both FSS approaches are subject to strong corrections to scaling induced by the frustration, their behavior is rather different, in particular as the ratio $\J2 = J_2/J_1$ is varied. As a consequence, an analysis of the scaling of partition function zeros turns out to be a useful complement to a more traditional FSS analysis. For the cumulant method, we also study the convergence as a function of cumulant order, providing suggestions for practical implementations. The scaling of the zeros convincingly shows that the system remains in the Ising universality class for $\J2$ as low as $-0.22$, where results from traditional FSS using the same simulation data are less conclusive. The approach hence provides a valuable additional tool for mapping out the phase diagram of models afflicted by strong corrections to scaling.}


\DeclareMathOperator{\J2}{\mathcal{R}}
\def\E{\Sigma}
\def\dlm{\mathrm{dln}|m|}
\def\sg{(\mathrm{d})}
\def\F{\mathrm{F}}
\def\LY{\mathrm{LY}}
\def\PA{\mathrm{PA}}

\def\bcA{1.02351(22)}
\def\bcB{2.65740(68)}
\def\bcC{3.25722(84)}
\def\bcD{4.27017(48)}

\PassOptionsToPackage{hyphens}{url}
\usepackage{xurl}
\begin{document}
\sloppy

\hypersetup{
  breaklinks=true,   
  colorlinks=true,   
  pdfusetitle=true,  
}
\section{Introduction}

Frustrated systems~\cite{Diep2012} are characterized by interactions that cannot all be satisfied simultaneously. The resulting internal competition leads to quite interesting critical behavior such as reentrant phases~\cite{Azaria1987} and non-zero ground state entropy~\cite{Espriu2004,Mueller2014}. 
One of the most well-studied systems in this class is the Ising model with competing first and second-neighbor interactions on the square lattice~\cite{nightingale:77,Jin2012,Kalz2011,Kalz2012,Yoshiyama2023}. Noting that in the presence of frustration the lattice geometry is of fundamental importance for the occurrence and symmetry of ordered phases, it is somewhat surprising that much less is known about the analogous system on the honeycomb lattice, which only recently started attracting some attention~\cite{Bobak2016,Acevedo2021,Zukovic2021,Schmidt2021}. Depending on the ratio $\J2$ of the next-nearest-neighbor and nearest-neighbor interaction strengths, the system has a ferromagnetic ground state for $\J2>-1/4$ or a largely degenerate ground state of known energy for $\J2 < -1/4$~\cite{Bobak2016}. The understanding of the critical properties for $\J2\lesssim-0.2$ remains limited. Here we focus on $\J2>-1/4$ for which the system remains ferromagnetic at zero temperature.

Metastable states are a common feature in frustrated systems and their presence is a challenge for standard simulation techniques since runs get trapped in local minima. Particularly difficult are systems with rugged free-energy landscapes~\cite{Janke2008}. One contender among generalized-ensemble simulation techniques suitable for such problems is population annealing (PA)~\cite{Iba2001,Hukushima2003} that has recently shown its versatility in a range of applications~\cite{Machta2010,Wang2015,Christiansen2019a}. PA is particularly well suited for studying such systems with many competing minima, as the large number of replicas allows sampling many local minima simultaneously. This is in contrast to its one-replica counterpart, equilibrium simulated annealing~\cite{Rose2019}, which is much more likely to get trapped in a single local minimum and hence to fail to sample the equilibrium distribution.

Systems with competing interactions often are subject to strong corrections to scaling (commonly of unknown shape). For the above-mentioned frustrated Ising model on the square lattice, for example, which has been discussed since the 1970s~\cite{nightingale:77}, the study of the tricritical point required simulations of rather large system sizes, but notwithstanding this effort certain aspects remain unclear~\cite{Kalz2011,Jin2012,Kalz2012,Yoshiyama2023}.

An alternate approach for studying phase transitions revolves around considering zeros of the partition function in the complex external field and temperature planes, based on the pioneering work by Lee and Yang~\cite{Yang1952,Lee1952}, and by Fisher~\cite{fisher1965nature}. 
Using these zeros allows one to distinguish between first- and second-order transitions as well as to extract estimates of their strengths~\cite{Janke2001,Janke2002b,Janke2002a,Janke2002}, and
to examine the peculiar properties of a model with special boundary
conditions~\cite{Janke2002c,Janke2002d}.
The cumulative density of zeros and their impact angle onto the real
temperature axis encode the strength of higher-order phase transitions~\cite{Janke2003,Janke2004,Janke2005,Kenna2005,Janke2006}.
This can also be used as a medium for deriving scaling relations among
logarithmic correction exponents~\cite{Kenna2006,Kenna2006a}. For a recent discussion
along an alternative route, see Ref.~\cite{Moueddene2024a}.

For non-frustrated systems the scaling of partition function zeros has been shown to yield quite accurate results for critical exponents even when using rather small system sizes~\cite{Deger2019,Moueddene2024}, suggesting that this might also be the case for more complex systems. Previous work studying partition function zeros for frustrated spin systems includes Refs.~\cite{Monroe2007,Kim2015,Sarkanych2015,Kim2021}.

Despite some experimental realizations of zeros in complex external magnetic fields~\cite{Peng2015,Ananikian2015,flaschner:18}, the most established approach of studying complex partition function zeros requires an accurate estimate of the full density of states, which is difficult to obtain both in experimental as well as in simulational setups (but see Ref.~\cite{macedo:23}). For simulations, this commonly results in limiting the maximum system size that can be studied.
More recently, an alternative method for obtaining the leading partition function zero based on cumulants of the energy and magnetization has gained some traction~\cite{Flindt2013}, thus enabling determinations of partition-function zeros from more easily accessible observables in simulations.
We use this method to perform finite-size-scaling (FSS) of the zeros for system sizes exceeding those for which we could determine the full density of states.

For the frustrated Ising model on the honeycomb lattice, initial work using effective field theory (EFT)~\cite{Bobak2016} suggested the existence of a tricritical point near $\J2\approx -0.1$. This was later challenged through a Monte Carlo study~\cite{Zukovic2021}, showing that the system remains within the Ising universality class at least down to $\J2=-0.2$. Cluster mean-field theory~\cite{Schmidt2021} suggests that the transition may remain of second order down to $\J2 = - 1/4$, but this has not been verified in the actual model. With this paper, we aim at a better understanding of the critical properties of this system, particularly close to the special point $\J2 = - 1/4$ at which the critical temperature vanishes. To this end we consider the scaling of the partition function zeros.

\section{Materials and Methods}

\subsection{Model}
We study the well-known two-dimensional Ising model, but placed on the honeycomb lattice and equipped with competing nearest- and next-nearest-neighbor interactions, resulting in the Hamiltonian
\begin{equation}
    \mathcal{H} = -J_1 \sum_{\langle ij \rangle} \sigma_i \sigma_j - J_2 \sum_{[ ik ]} \sigma_i \sigma_k - h \sum_i \sigma_i \equiv -J_1 \E_1 - J_2 \E_2 - h M,
\end{equation}
where $\langle ij \rangle$ and $[ ik ]$ denote sums over nearest neighbors and next-nearest neighbors, respectively, $J_1>0$ is the ferromagnetic nearest-neighbor interaction strength, $J_2 < 0$ is the competing antiferromagnetic next-nearest-neighbor interaction strength, and $h$ the magnetic field\footnote{Note that all simulations were carried out in the absence of an external magnetic field. We include the magnetic term here as it is necessary for the discussion of the Lee-Yang zeros.}. $\E_1$ and $\E_2$ refer to the sums over nearest- and next-nearest-neighbor interactions, respectively, and $M$ is the (total) magnetization. The quantity relevant for the nature of the ordered phase and the transition is the ratio $\mathcal{R}=J_2/J_1$ of couplings. Here we only consider the case of $\mathcal{R} \in (-1/4,0]$, where the ground state is ferromagnetic~\cite{Bobak2016}.
We study systems of linear lattice size $L$, which due to the two-atom basis of the honeycomb lattice contain $N=2 L^2$ spins.
\subsection{Partition function zeros}
In terms of the density of states $\Omega(\E_1,\E_2,M)$, the partition function at inverse temperature~$\beta$ and external magnetic field $h$ is given by
\begin{equation}
	\mathcal{Z}(\beta,h) = \sum_{\E_1} \sum_{\E_2} \sum_{M} \Omega(\E_1,\E_2,M) e^{\beta J_1 \E_1 + \beta J_2 \E_2 + \beta h M}.\label{eq:partFunc}
\end{equation}
We choose both $J_1$ and $J_2$ to be integers, such that for $h=0$ the partition function is a polynomial in $x=e^{-\beta}$. For $h \neq 0$, on the other hand, the partition function may be written as a polynomial in $e^{2 \beta h}$ for arbitrary choices of $J_1$ and $J_2$.
The complex inverse temperatures $\{\beta_k\}$ solving the equation $\mathcal{Z}(\beta=\beta_k,h=0)=0$, i.e., in the absence of an external magnetic field $h$,  are called Fisher zeros.
Once calculated, the Fisher zeros will be studied as a function of $z = x^{J_1}$ to allow for better comparability of the results for different values of $\J2=J_2/J_1$.
Note that, as is well known, different variables yield very different visual impressions for the locations of zeros (see, for example, Appendix~\ref{sec:FisherZeros_J2_0}).
The pair of zeros closest to the positive real axis approaches $(\beta_c,0)=\beta_c + 0i$ as $L\rightarrow\infty$. The real and imaginary parts of these leading zeros $\beta_0$ usually scale as~\cite{Itzykson1983}
\begin{subequations}
	\begin{equation}
		\Re(\beta_0) - \beta_c \propto L^{-y_t},
	\end{equation}
	and
	\begin{equation}
		\Im(\beta_0) \propto L^{-y_t},
	\end{equation}
\end{subequations}
respectively, with $y_t$ being the renormalization-group (RG) eigenvalue related to the temperature variable, which is connected to the critical exponent of the correlation length by $y_t=1/\nu$.

The complex magnetic fields $\{h_k\}$ that solve the equation $\mathcal{Z}(\beta,h=h_k)=0$ for some fixed $\beta$ are the so-called Lee-Yang zeros that lie on the unit circle $e^{2\beta h} = e^{i \varphi}$ for the Ising ferromagnet, implying that all solutions for $h_k$ are purely imaginary~\cite{Lee1952,Yang1952}.
This circle theorem has been extended to many more models~\cite{Bena2005}, but it is not universally valid, see, e.g., Refs.~\cite{Froehlich2012,Krasnytska2015}. For the Ising model with competing interactions placed on a square lattice the circle law was found to apply in the regime with ferromagnetic ground state~\cite{Katsura1971}.
As $L \rightarrow \infty$, the Lee-Yang zeros closest to the positive real axis approach zero. At the inverse critical temperature $\beta_c$ the imaginary part of the leading zeros $h_0$ scales as~\cite{Itzykson1983}
\begin{equation}
	\Im(h_0) \propto L^{-y_h},\label{eq:scalingLY}
\end{equation}
with $y_h$ being the RG eigenvalue related to the external magnetic field, which is connected to the standard critical exponents by $y_h=(\beta+\gamma)/\nu$.

To numerically estimate partition function zeros, writing $\beta=\Re(\beta)+i\Im(\beta)$ one notes that for zero field, $h=0$, the Fisher zeros of $\mathcal{Z}(\beta,h=0)$ are identical to the zeros of 
\begin{subequations}
	\begin{equation}
		Z(\Re(\beta), \Im(\beta)) \equiv \frac{\mathcal{Z}(\Re(\beta)+i\Im(\beta),h=0)}{\mathcal{Z}(\Re(\beta),h=0)} = \langle \cos ( \Im(\beta) \mathcal{H} ) \rangle_{\Re(\beta),h=0} - i \langle \sin ( \Im(\beta) \mathcal{H} ) \rangle_{\Re(\beta),h=0}, \label{eq:complexPartFunc}
	\end{equation}
	and likewise the Lee-Yang zeros for fixed (real) $\beta$ might be extracted from
	\begin{equation}
		Z_\beta(\Re(h), \Im(h)) \equiv \frac{\mathcal{Z}(\beta,\Re(h)+i \Im(h))}{\mathcal{Z}(\beta,\Re(h))} = \langle \cos ( \Im(h) M ) \rangle_{\beta,\Re(h)} - i \langle \sin ( \Im(h) M ) \rangle_{\beta,\Re(h)}. \label{eq:complexPartFuncH}
	\end{equation}
\end{subequations}

While the evaluation of \eqref{eq:complexPartFunc} and \eqref{eq:complexPartFuncH} and a systematic search for zeros requires the availability of the density of states $\Omega$ for reweighting~\cite{Kenna1991,Kenna1994,Hong2020,Moueddene2024}, more recently a computationally lighter method based on cumulants of thermodynamic observables $\Phi(q)$, i.e.,
\begin{equation}
    \langle\!\langle \Phi^n (q) \rangle\!\rangle = \frac{\partial^n}{\partial q^n} \ln\mathcal{Z}(q) \label{eq:cumulantsDef}
\end{equation}
has been suggested~\cite{Flindt2013,Deger2018,Deger2019,Deger2020,Deger2020a}. Here, the notation $\Phi(q)$ refers to the thermodynamic observables $\mathcal{H}$ and $M$ as a function of their control parameters $q=-\beta$ and $q=\beta h$, respectively, unifying the discussion of Fisher and Lee-Yang zeros.
The method relies on the fact that the partition function can be factorized in a regular (non-zero) part $\tilde{\mathcal{Z}}(q)=\mathcal{Z}(0) e^{c q}$ for some constant $c$ and the product of its complex roots in $q$, i.e.,
\begin{equation}
    \mathcal{Z}(q) = \tilde{\mathcal{Z}}(q) \prod_k (1 - q / q_k).\label{eq:partFuncFactorized}
\end{equation}
Note that by Eq.~\eqref{eq:partFunc} $\mathcal{Z}(q)$ is real for real $q$, and that thus the roots $q_k$ in \eqref{eq:partFuncFactorized} appear in complex conjugate pairs.
Plugging (\ref{eq:partFuncFactorized}) into (\ref{eq:cumulantsDef}) yields the expression
\begin{equation}
     \langle\!\langle \Phi^n (q) \rangle\!\rangle = - \sum_k \frac{(n-1)!}{{(q_k-q)}^n}, \quad n > 1 \label{eq:phiCum}
\end{equation}
for the cumulants. The key point of the method is that the contribution of non-leading zeros in the expression above is suppressed by powers of $n$ for the $n$-th order cumulant. Thus, one neglects the non-leading zeros, which allows the calculation of the leading partition function zeros using only the first few cumulants of the energy and magnetization, $\langle\!\langle \mathcal{H}^{n}\rangle\!\rangle$ and $\langle\!\langle M^{n}\rangle\!\rangle$, and hence does not require knowledge of $\Omega$.
Cumulants can be calculated from the central moments, $\langle \Phi^n \rangle_c = \left\langle{(\Phi-\langle \Phi \rangle)}^n\right\rangle$. The first four cumulants are given by
\begin{equation}
	\langle\!\langle \Phi \rangle\!\rangle = \langle \Phi \rangle, \quad \langle\!\langle \Phi^2 \rangle\!\rangle = \langle \Phi^2 \rangle_c,  \quad \langle\!\langle \Phi^3 \rangle\!\rangle = \langle \Phi^3 \rangle_c,  \quad \langle\!\langle \Phi^4 \rangle\!\rangle = \langle \Phi^4 \rangle_c - 3  \left(\langle \Phi^2 \rangle_c\right)^2.
\end{equation}
Relations for higher-order cumulants can be found using computer algebra systems.\footnote{We use the \textsc{Mathematica} function \texttt{MomentConvert} to obtain the relations up to the 20-th cumulant. The first ten cumulants are listed in Ref.~\cite{Janke1997}.}
Within this framework, the leading zeros $q_0$ can be extracted in a vector-matrix notation from~\cite{Flindt2013}
	\begin{equation}
		\label{eq:cumulantMethod}
		\begin{pmatrix}
			2~\Re(q_{0}-q)\\
			|q_{0}-q|^2
		\end{pmatrix}\approx\begin{pmatrix}
			1& \displaystyle-\frac{{\mu}_{n}^{(+)}}{n}\\
			1& \displaystyle-\frac{{\mu}_{n+1}^{(+)}}{n+1}
		\end{pmatrix}^{-1}
		\begin{pmatrix}
			(n-1) {\mu}_{n}^{(-)}\\
			n \ {\mu}_{n+1}^{(-)}
		\end{pmatrix},
	\end{equation}
    where $n$ is the approximation order, and $\mu_{n}^{(\pm)}$ denotes the ratio of two cumulants of consecutive orders, i.e., $\mu_{n}^{(\pm)} \equiv \langle\!\langle \Phi^{n\pm 1}\rangle\!\rangle / \langle\!\langle \Phi^{n}\rangle\!\rangle$. $(\Phi,q) = (E,-\beta)$ corresponds to Fisher zeros, and $(\Phi,q) = (M,\beta h)$ to Lee-Yang zeros.
    Since odd cumulants of $M$ vanish for $h=0$, setting $n=2k$ the expression for the Lee-Yang zeros simplifies to~\cite{Deger2020a,Moueddene2024}
    \begin{equation}
        \label{eq:cumulantMethodLY}
        \Im (h_0) \approx \pm \frac 1 \beta \sqrt{2k (2k+1) \left|\frac{\langle\!\langle M^{2k}(0)\rangle\!\rangle}{\langle\!\langle M^{2(k+1)}(0)\rangle\!\rangle}\right|}.
    \end{equation}

\subsection{Population annealing}

Population annealing (PA)~\cite{Iba2001,Hukushima2003} is a simulation scheme designed for parallel calculations for systems with complex free-energy landscapes in which the state space is sampled by a population of replicas that are cooled down collectively. 
It consists of alternating resampling and spin-update steps. 
As the temperature is lowered, the population is resampled according to the Boltzmann distribution at the new temperature~\cite{Gessert2023}.
Spin updates then help to equilibrate the replicas at this temperature, and to increase the diversity of the population~\cite{Weigel2021}.
The implementation of PA as used here can be summarized as follows:

\begin{enumerate}
	\item Initialize the population by drawing $R_0=R$ random spin configurations corresponding to the initial inverse temperature $\beta_0^\PA=0$.
	
	\item Set the iteration counter $i \leftarrow 0$.
	
	\item Determine the next inverse temperature $\beta_{i+1}^\PA$ such that the energy histogram overlap between $\beta_i^\PA$ and $\beta_{i+1}^\PA$, given by~\cite{Barash2017}
	\begin{equation}
		\alpha (\beta_{i}^\PA,\beta_{i+1}^\PA) = \frac{1}{R_i} \sum_{j=1}^{R_i} \min \left(1, \frac{R e^{-(\beta_{i+1}^\PA - \beta_i^\PA) E_j}}{\sum_{k=1}^{R_i} e^{-(\beta_{i+1}^\PA - \beta_i^\PA) E_k}}\right),
	\end{equation}
	is approximately equal to the target value of $\alpha^{*}$, where $E_j$ refers to the current energy of replica $j$.
	
	\item Increment $i$ by 1.
	
	\item Resample the replicas according to their relative Boltzmann weights, that is, make on average
	\begin{equation}
		\tau(E_j) = \frac{R e^{-(\beta_{i+1}^\PA - \beta_i^\PA) E_j}}{\sum_{k=1}^{R_i} e^{-(\beta_{i+1}^\PA - \beta_i^\PA) E_k}}
	\end{equation}
	copies of replica $j$ with energy $E_j$.
	
	\item Carry out Metropolis updates on the replicas until the effective population size $R_\text{eff}$ (see \cite{Weigel2021} for definition and discussion) exceeds the threshold value of $R^{*}=\rho^{*} R$.
	
	\item Calculate estimates for observables $\mathcal{O}$ as the population averages $\sum_j \mathcal{O}_j / R_i$, where $\mathcal{O}_j$ is the value of the observable for the $j$-th replica.\footnote{Note that we calculate central moments of the energy directly during the simulation after calculating the average energy, because using raw moments to calculate higher order central moments and cumulants leads to a complete loss of numeric precision in the latter.}
	
	\item Unless the lowest temperature of interest is reached, go to step 3.
\end{enumerate}

Some comments are in order at this point.
The above implementation contains numerous parameters of relevance to the performance of the algorithm, namely the population size $R$, the target energy distribution overlap $\alpha^{*}$, and the sweep schedule given by the threshold value for the relative effective population size $\rho^{*}$.
These have been (resp.\ will be) discussed elsewhere~\cite{Weigel2021,Gessert2023,Gessert2024} and here we choose $R=20\,000$ throughout, $\alpha^{*}$ is set to at least $80\%$, in some cases $90\%$, and $\rho^{*} \geq 90\%$.
There are in fact many possibilities of how the resampling step can be realized. Here, we use the so-called nearest-integer resampling which was shown to lead to optimal results in many scenarios~\cite{Gessert2023}. For the spin updates we employ the Metropolis method implemented on GPU drawing on a domain decomposition~\cite{Barash2017} into four sublattices, adapting and extending the publicly available code of Ref.~\cite{Barash2017}.
For small system sizes, we obtain an estimate for the density of states by using multi-histogram reweighting, also implemented in the source code of Ref.~\cite{Barash2017}.

\section{Results}
\subsection{Solving for all Fisher zeros in the complex temperature plane}

For finite systems, the partition function for $h=0$ is a polynomial of finite order in $x=e^{-\beta}$ (for suitable choices of $J_1$ and $J_2$). In principle, once the density of states $\Omega(E)$ is known one can solve for all partition function zeros numerically.
Due to the presence of both $J_1 \E_1$ and $J_2 \E_2$ in the Hamiltonian, there are more than the usual $\mathcal{O}(N)$ distinct energy levels, namely up to $\mathcal{O}(N^2)$ levels, where $N=2 L^2$ is the number of spins in the honeycomb system.
Thus, it is only feasible to calculate all partition function zeros for very small system sizes.

\begin{figure}[t]
    \includegraphics{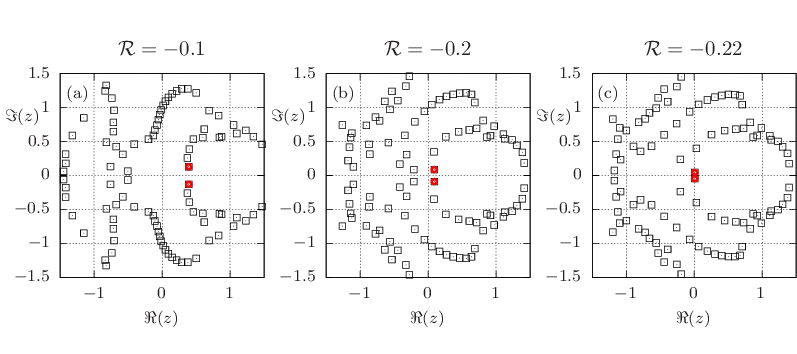}
    \caption{Fisher zeros in the (transformed) complex temperature $z = e^{-\beta J_1}$ plane for $L=4$ and (a) $\J2=-0.1$, (b) $\J2=-0.2$, and (c) $\J2=-0.22$. The red circles highlight the Fisher zeros closest to the positive real axis.\label{fig:allZeros_L4}}
\end{figure}

This is illustrated in Figure~\ref{fig:allZeros_L4}, where all\footnote{Note that \textsc{Mathematica} fails to find some zeros close to the real axis which becomes apparent by some zeros visible for $\J2=-0.1$ and $-0.2$ being absent for $\J2=-0.22$.} Fisher zeros of a 32-spin system ($L=4$) with periodic boundary conditions for different values of $\J2$ are depicted.
To compute the density of states we used exact enumeration of all possible $2^{32}$ spin configurations. In the absence of next-nearest-neighbor interactions (i.e., for the standard ferromagnetic Ising case) the model can be solved analytically for $L\rightarrow\infty$, see Appendix~\ref{sec:FisherZeros_J2_0} for a comparison of the partition function zeros for $L=4$ with the exact solution in the thermodynamic limit. For $\J2=0$, the partition function is an even polynomial in $z=x^{J_1}$ and thus invariant under sign inversion of $z$. This symmetry is broken by the introduction of the next-nearest-neighbor interactions, which is reflected in the asymmetry with respect to the ordinate axis in Figure~\ref{fig:allZeros_L4} that becomes stronger with increasing $|\J2|$ (left to right).
Also note that the number of zeros for the finite system size is much larger than for $\J2= 0$. This is due to a smaller number of distinct energy levels for $\J2=0$, or in other words, the larger degeneracy of the individual energy levels.
For $L=4$, the Fisher zeros at $z = \pm i$ for $\J2=0$ each split into 16 distinct ones when $|\J2|$ is increased, and spread out as $|\J2|$ is increased further.
For $\J2 = -0.1$ this is well seen through the dense set of zeros near $z = \pm i$. 
For a better visual impression of how the Fisher zeros move as $\J2$ is changed, we refer to the Supplementary Materials which contain an animation (video) illustrating the motion of Fisher zeros as $\J2$ is varied from $0$ to $-1$, illustrating the splitting of the zero located at $z = \pm i$ for $\J2 = 0$ very clearly.

From earlier studies~\cite{Bobak2016,Zukovic2021}, it is known that the critical temperature in this model approaches zero as $\J2$ goes to $-1/4$. This is reflected in Figure~\ref{fig:allZeros_L4} by the leading Fisher zero (red circle) moving closer to the origin in this limit.
Due to the small system size, the leading Fisher zero for $\J2=-0.22$ is very close to the imaginary axis. In fact, for $\J2 \lesssim -0.222$, the imaginary part of the leading Fisher zero in the $\beta$-plane, $\Im(\beta_0)$, exceeds $\pi/2$. Thus, in the $z$-plane the zero may lie in the region of negative real values for smaller values of $\J2$. As $\J2$ approaches $-1/4$, both the real and imaginary part of $\beta_0$ go to infinity. Thus, in the $z$-plane, the Fisher zero rotates around the origin as $\J2$ goes to $-1/4$. Since the imaginary part of $\beta_0$ vanishes with increasing $L$, we understand this effect as a peculiar finite-size effect and expect it not to be relevant in the thermodynamic limit.

\subsection{Determining the leading Fisher zero directly and by the cumulant method}\label{sec:FisherZerosGraphically}
In the following we verify the efficacy of the cumulant method developed by Flindt and Garrahan~\cite{Flindt2013} by comparing its results to the estimates from reweighting for system sizes for which we obtained the density of states. As we were unable to calculate all partition function zeros for $L > 4$, we fall back to obtaining the value of the leading Fisher zero used for comparison by numerically reweighting $Z(\Re(\beta),\Im(\beta))$. 
Except for $L=4$ where exact enumeration was used, we estimate the density of states\footnote{Note that we only measured the two-dimensional (energetic) density of states. Therefore, we do not have access to magnetic quantities and the Lee-Yang zeros.} $\Omega(\E_1,\E_2)$ by using multi-histogram reweighting (MHR) of our PA data~\cite{Barash2017}.

\begin{figure}[t]
    \includegraphics{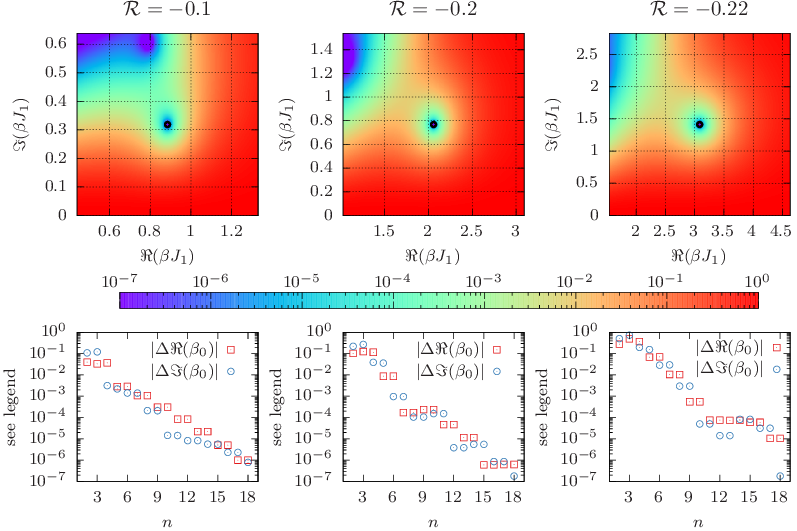}
    \caption{The top row shows heat maps of $|Z(\Re(\beta),\Im(\beta))|$ for complex $\beta$ as obtained from Equation~(\ref{eq:complexPartFunc}) for $\J2\in\{-0.1,-0.2,-0.22\}$ calculated by exact enumeration for $L=4$. The leading Fisher zero calculated by the direct method is denoted by an open black circle. The bottom row shows the differences $\Delta\Re(\beta_0)$ and $\Delta\Im(\beta_0)$ (see text for definition), comparing the direct method with the cumulant method.\label{fig:directVsCumulant_L4}}
\end{figure}

The top row of Figure~\ref{fig:directVsCumulant_L4} shows the absolute value of the partition function $Z(\Re(\beta),\Im(\beta))$ in the complex $\beta$-plane as obtained from Equation~(\ref{eq:complexPartFunc}) for different values of $\J2$, zoomed-in and centered around the leading Fisher zero for $L=4$ with positive imaginary part (using the same data as above). The open black circles denote the respective Fisher zero $\beta^{\sg}_0$ found using the Levenberg-Marquardt (LM) algorithm~\cite{Levenberg1944,Marquardt1963} with an initial guess for $\beta^{\sg}_0$ close to the root.\footnote{We use \textsc{SciPy}'s \texttt{optimize.root} function to find the zeros with the parameter \texttt{method='lm'} (Levenberg-Marquardt algorithm).} In the following, we will refer to this approach using the LM algorithm as the ``direct method''. Note the added superscript $\sg$ for the value of the leading Fisher zero obtained with this method. A commonly used alternative approach is to use one-dimensional root finding to determine the zeros of the real and imaginary parts of Equations~(\ref{eq:complexPartFunc}) or (\ref{eq:complexPartFuncH}) independently over a range of complex $\beta$'s or $h$'s, respectively, and then to find their intersection, as was done in Refs.~\cite{Kenna1991,Kenna1994,Hong2020,Moueddene2024}.

For the cumulant method, as the cumulant order $n$ goes to infinity, when the cumulants are evaluated at $\beta=\Re(\beta^{\sg}_0)$ the estimate for $\Re(\beta-\beta_0)$ goes to zero and the approximation of $|\beta-\beta_0|^2$ approaches $\Im{(\beta^{\sg}_0)}^2$.\footnote{In principle, one can use any simulation point $\beta$ and obtain estimates for the location of the Fisher zero. However, most precise results are found for $\beta \approx \Re(\beta_0)$.} Thus, we can consider $\Delta\Re(\beta_0)\equiv\Re(\beta-\beta_0)|_{n,\beta=\Re(\beta^{\sg}_0)}$ and $\Delta\Im(\beta_0)\equiv\sqrt{|\beta-\beta_0|^2}\big|_{n,\beta=\Re(\beta^{\sg}_0)} - \Im(\beta^{\sg}_0)$ to probe the rate of convergence in $n$.
The bottom panel shows the absolute values $|\Delta\Re(\beta_0)|$ and $|\Delta\Im(\beta_0)|$ of the differences as a function of $n$, which appear to decay exponentially. This observation is in line with the fact that the contributions of sub-leading partition function zeros are suppressed with power $n$, see Equation~(\ref{eq:phiCum}).

\begin{figure}[t]
    \includegraphics{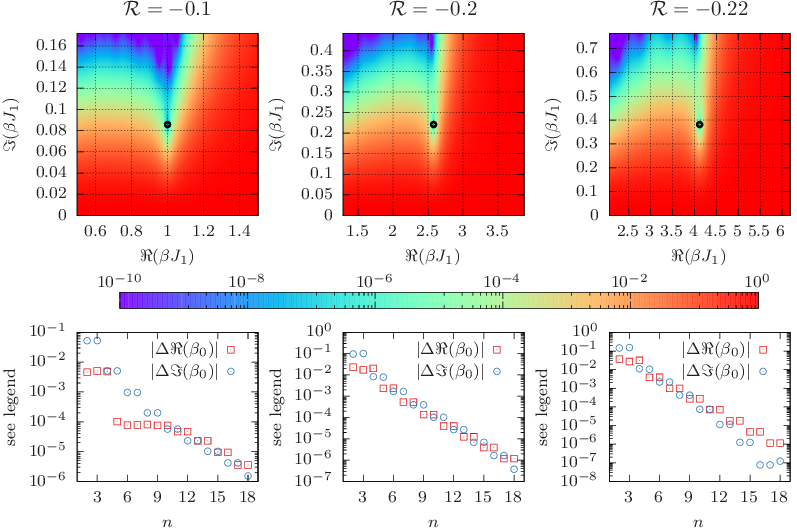}
    \caption{Same as Figure~\ref{fig:directVsCumulant_L4} for $L=16$, but using PA simulation data instead of exact enumeration\label{fig:directVsCumulant}.}
\end{figure}

Next, we repeat the same analysis with data for the density of states obtained by MHR of PA data with system size $L=16$. For each value of $\J2$ we carried out an independent simulation. Although reweighting in $J_2$ is in principle possible, the reweighting range is rather limited. 
The results of this exercise are shown in Figure~\ref{fig:directVsCumulant}, and they are found to be qualitatively quite similar to the previous case.
As is to be expected, the leading Fisher zero is closer to the real axis as compared to $L=4$.
Also here the cumulants are evaluated at the $\beta$ equal to the estimate for $\Re(\beta_0)$ from the direct method which is possible thanks to the estimate of the density of states $\Omega$ from MHR.
In particular, the exponential decay of the differences between the cumulant and direct method is more clearly visible in this case.
Note that the bottom row only shows the systematic deviation of the two methods when using the same data for $\Omega(\Sigma_1,\Sigma_2)$ (subject to statistical errors), and not the actual error for $\beta_0$.
For an estimate of the statistical errors encountered in the simulation, see the error bars of Figure~\ref{fig:cumulantsDiffJ2_and_L}.
This demonstrates that the cumulant method is a viable replacement for the direct method whenever the statistical error exceeds the systematic deviation shown above (which it typically does). However, it does not say anything about the actual accuracy of the obtained results.
Also note that even on the logarithmic scale, the difference decays monotonously and no noise is visible even when using higher-order cumulants. This is because despite the fact that higher-order cumulants are noisy, the fluctuation of their ratios does not increase noticeably with $n$, which is due to the cross-correlation between the terms.%

\subsection{Determining the leading Lee-Yang zero directly and by the cumulant method}

We now turn to an analysis of the Lee-Yang zeros. We consider the partition function in the complex-field plane at our best estimate for the infinite-volume critical temperature for different values of $\J2$ (see Sec.~\ref{sec:fss} for details). Analogous to the Fisher zeros discussed in the previous section, Lee-Yang zeros are obtained using the direct method (utilizing the LM algorithm). As their calculation requires the density of states $\Omega(\E_1,\E_2,M)$, we only compute them for $L=4$. 

\begin{figure}[t]
    \includegraphics{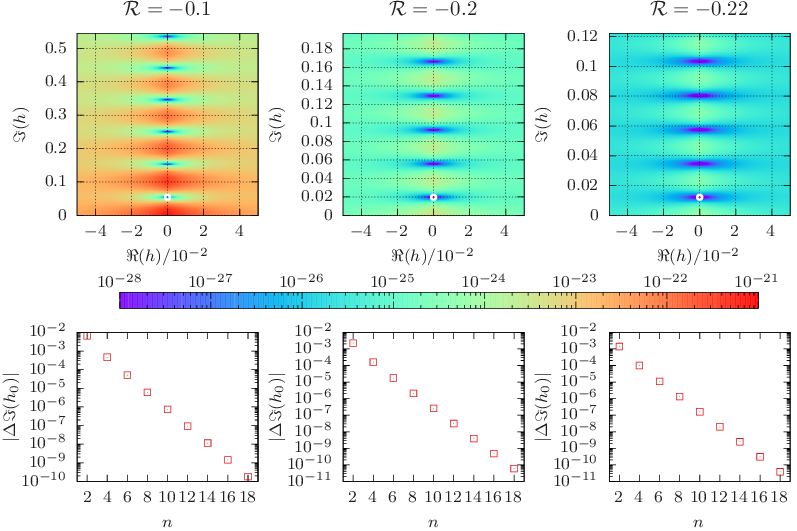}
    \caption{The top row shows $|Z_{\beta_c}(\Re(h),\Im(h))|$ for complex $h$ as obtained from Equation~(\ref{eq:complexPartFuncH}) for $\J2\in\{-0.1,-0.2,-0.22\}$ calculated by exact enumeration for $L=4$ at $\beta_c$. For $\beta_c$ we used our best estimate, see Section~\ref{sec:fss}. The leading Lee-Yang zero is denoted by an open white circle. The bottom row shows the difference in $\Im(h_0)$ for the approximation given by Equation~(\ref{eq:cumulantMethodLY}) of order $n$ and the directly obtained value.\label{fig:directVsCumulant_LY_L4}}
\end{figure}

Similar to Figure~\ref{fig:directVsCumulant_L4}, the top panel of Figure~\ref{fig:directVsCumulant_LY_L4} shows the absolute value of the partition function in the complex field plane at the infinite-volume inverse critical temperature $\beta_c$, i.e., $|Z_{\beta_c}(\Re(h),\Im(h))|$.
The found zeros are consistent with being purely imaginary, suggesting that the Lee-Yang circle theorem may also hold for the model considered here.\footnote{Note that we are unaware of any rigorous proof of the circle theorem for the model at hand.}
We again also considered the alternative approach provided by the cumulant method, and the bottom panel of Figure~\ref{fig:directVsCumulant_LY_L4} depicts the deviation of the estimate of the cumulant method from the zero determined via the direct method as a function of $n$. %
As was the case for the Fisher zeros, the deviation vanishes exponentially in $n$. 
Note that the range of the real part of the external magnetic field is the same in all three panels.
For smaller values of $\J2$, the minimum of $|Z_{\beta_c}(h)|$ becomes broader making it more difficult to find the root numerically.
The imaginary part of the leading Lee-Yang zero vanishes as $\J2$ approaches $-1/4$.

\subsection{Comparison of standard FSS and scaling of partition function zeros}\label{sec:fss}
One recently proposed advantage of using the partition function zeros to obtain critical exponents is that already rather small system sizes may yield quite accurate estimates for the exponents~\cite{Moueddene2024}.%
In the following we will present tables for different values of $\J2$ and different fit intervals from $L_{\min}$ to $L_{\max}$, thus clearly demonstrating for which ranges of system sizes reliable FSS fits can be performed.
We have considered the system sizes $L \in \{ 8, 16, 24, 32, 48, 64, 88\}$ and values of $\J2$ equal to $-0.1$, $-0.2$, $-0.21$, and $-0.22$. As the results for $\J2=-0.21$ are analogous to the other values of $\J2$, they are only included in the Supplementary Materials, and not presented in the main text.
For every system size $L$ and for every value of $\J2$ ten independent PA runs were carried out in order to improve statistics and to obtain error bars. The first run determined the temperature set for the remaining runs.\footnote{In principle, the energy histogram overlap defines the temperature set uniquely. However, the actually determined temperatures by PA are subject to statistical fluctuations. Thus, to avoid every realization having its own temperatures, the first run is used to fix the temperature set.} The same simulation data was used for the comparison of standard FSS and scaling of partition function zeros obtained via the cumulant method using up to the 20-th cumulant (corresponding to $n=18$ in Equation~(\ref{eq:cumulantMethod}) for Fisher zeros, and $k=9$ in Equation~(\ref{eq:cumulantMethodLY}) for Lee-Yang zeros). As shown above, we did not observe any loss in numeric precision using higher-order cumulants. Therefore, we use the largest cumulant that we measured; see also Appendix~\ref{app:convergenceCumulants}.

\subsubsection{Scaling of Fisher zeros}
Table~\ref{tab:fss_yt} summarizes the FSS results for $\J2 = -0.1$, $-0.2$, and  $-0.22$ using the leading Fisher zero as well as the logarithmic derivative of the magnetization,
\begin{equation}
	\dlm \equiv \frac{\partial}{\partial \beta} \ln \langle |M| \rangle = \frac{\langle |M| \mathcal{H} \rangle}{\langle |M| \rangle}-\langle\mathcal{H}\rangle,
\end{equation}
whose maximum follows the FSS relation $\dlm_{\max}(L) \propto L^{y_t}$~\cite{ferrenberg:91a}.
As the precise inverse temperature at which the $\Re(\beta-\beta_0)$ is zero is not in general contained in the annealing schedule of PA, we first determine the zero crossing in the estimate for $\Re(\beta-\beta_0)$ with positive slope closest to the location of the local maximum of the second energy cumulant (ignoring the $\beta=0$ maximum). Next, we determine the $\beta$ for which the linear interpolation of $\Re(\beta_0)-\beta$ of Equation~\eqref{eq:cumulantMethod} between the two closest inverse-temperatures is zero. $\Im(\beta_0)$ is also obtained through linear interpolation to the same $\beta$.\footnote{We found this approach to be more stable than to evaluate $\Re(\beta_0)$ and $\Im(\beta_0)$ directly using Equation~\eqref{eq:cumulantMethod}.}
This is subsequently used for FSS.
Details from all fits yielding estimates for $y_t$ can be found in Section~2.1 of the Supplementary Materials. %
The expected exponent is the Onsager value $y_t=1$. The estimates $y_t^{\F}$ and $y_t^{\dlm}$ for $y_t$ resulting from both methods are close to the expected value for the considered ranges of system sizes for all $\J2$, albeit not always within error bars.

For $\J2=-0.1$ it comes as a surprise that the values from standard FSS using $\dlm$ are closer to $1$ and appear to have weaker corrections to scaling. 
Specifically, the value obtained for $y_t^{\dlm}$ is within 0.5\% of the expected value for all fit ranges, whereas the value for $y_t^{\F}$ differs by as much as 3\% when using the fitting range $8-24$ for $L$.
While when $L_\text{min}=8$ the value for $y_t$ using the leading Fisher zero is far below 1, it is consistent with $1$ for larger $L_\text{min}$, suggestive of the differing value for smaller system sizes being due to stronger corrections to scaling.

In contrast, for $\J2 = -0.2$ for almost any (fixed) fitting range the two estimates are compatible with each other within error bars, and the error bars are comparable in magnitude. Most of the $y_t$ estimates fall well below the expected value $y_t = 1$, and are increasing with both $L_{\min}$ and $L_{\max}$, being compatible with $1$ only for fitting ranges limited to the largest system sizes studied here. This effective variation of the exponent with $L$ is also reflected in the poor quality of fit: In most ranges $[L_{\min},L_{\max}]$ the $Q$ value\footnote{The $Q$ value refers to the probability to draw a $\chi^2$ from the $\chi^2$ distribution that is even larger than the value calculated from the fit. Unusually small numbers for $Q$ correspond to poor quality of fit.} falls below $0.1$~\cite{Press2007}.

\begin{table}[H]
    \caption{Estimates for $y_t$ from FSS fits using the imaginary part of the leading Fisher zero as well as $\dlm$ for different $L_{\min}$ and $L_{\max}$, and using $\J2 = -0.1$, $-0.2$, and $-0.22$.}\label{tab:fss_yt}
	\begin{adjustwidth}{-\extralength}{0cm}
    \begin{tabularx}{\fulllength}{llllll}

\toprule

$\mathbf{L_{\max}}$ & 24 & 32 & 48 & 64 & 88 \\

$\mathbf{L_{\min}}$ &  &  &  &  &  \\

\midrule
&  &  &$\J2 = -0.1$  &  &  \\
\midrule

8 & \makecell[l]{$y_t^{\F} = 0.9668(61)$\\$y_t^{\dlm} = 0.9957(24)$} & \makecell[l]{$y_t^{\F} = 0.9667(48)$\\$y_t^{\dlm} = 0.9943(17)$} & \makecell[l]{$y_t^{\F} = 0.9796(31)^{\dagger}$\\$y_t^{\dlm} = 0.9941(15)$} & \makecell[l]{$y_t^{\F} = 0.9817(29)^{\dagger}$\\$y_t^{\dlm} = 0.9939(13)$} & \makecell[l]{$y_t^{\F} = 0.9829(29)^{\dagger}$\\$y_t^{\dlm} = 0.9941(11)$} \\

16 & -- & \makecell[l]{$y_t^{\F} = 0.972(15)$\\$y_t^{\dlm} = 0.9867(96)$} & \makecell[l]{$y_t^{\F} = 1.0001(79)^{\dagger}$\\$y_t^{\dlm} = 0.9908(51)$} & \makecell[l]{$y_t^{\F} = 1.0026(70)$\\$y_t^{\dlm} = 0.9916(35)$} & \makecell[l]{$y_t^{\F} = 1.0052(68)^{\dagger}$\\$y_t^{\dlm} = 0.9932(23)$} \\

24 & -- & -- & \makecell[l]{$y_t^{\F} = 1.017(13)$\\$y_t^{\dlm} = 0.9908(51)$} & \makecell[l]{$y_t^{\F} = 1.018(11)$\\$y_t^{\dlm} = 0.9916(35)$} & \makecell[l]{$y_t^{\F} = 1.022(11)$\\$y_t^{\dlm} = 0.9933(23)$} \\

32 & -- & -- & -- & \makecell[l]{$y_t^{\F} = 1.042(21)$\\$y_t^{\dlm} = 0.9935(50)$} & \makecell[l]{$y_t^{\F} = 1.049(19)$\\$y_t^{\dlm} = 0.9948(31)$} \\

48 & -- & -- & -- & -- & \makecell[l]{$y_t^{\F} = 1.048(38)$\\$y_t^{\dlm} = 0.9959(65)$} \\

\midrule
&  &  &$\J2 = -0.2$  &  &  \\
\midrule

8 & \makecell[l]{$y_t^{\F} = 0.9215(35)$\\$y_t^{\dlm} = 0.9211(27)^{\dagger}$} & \makecell[l]{$y_t^{\F} = 0.9263(31)^{\dagger}$\\$y_t^{\dlm} = 0.9284(18)^{\dagger}$} & \makecell[l]{$y_t^{\F} = 0.9369(26)^{\dagger}$\\$y_t^{\dlm} = 0.9344(15)^{\dagger}$} & \makecell[l]{$y_t^{\F} = 0.9400(23)^{\dagger}$\\$y_t^{\dlm} = 0.9380(14)^{\dagger}$} & \makecell[l]{$y_t^{\F} = 0.9453(21)^{\dagger}$\\$y_t^{\dlm} = 0.9484(10)^{\dagger}$} \\

16 & -- & \makecell[l]{$y_t^{\F} = 0.9500(86)$\\$y_t^{\dlm} = 0.9493(47)$} & \makecell[l]{$y_t^{\F} = 0.9663(54)^{\dagger}$\\$y_t^{\dlm} = 0.9581(34)^{\dagger}$} & \makecell[l]{$y_t^{\F} = 0.9639(42)^{\dagger}$\\$y_t^{\dlm} = 0.9625(29)^{\dagger}$} & \makecell[l]{$y_t^{\F} = 0.9702(37)^{\dagger}$\\$y_t^{\dlm} = 0.9707(19)^{\dagger}$} \\

24 & -- & -- & \makecell[l]{$y_t^{\F} = 0.994(12)$\\$y_t^{\dlm} = 0.9694(64)^{\dagger}$} & \makecell[l]{$y_t^{\F} = 0.9780(83)$\\$y_t^{\dlm} = 0.9737(48)$} & \makecell[l]{$y_t^{\F} = 0.9872(68)^{\dagger}$\\$y_t^{\dlm} = 0.9788(26)$} \\

32 & -- & -- & -- & \makecell[l]{$y_t^{\F} = 0.968(15)^{\dagger}$\\$y_t^{\dlm} = 0.9829(65)$} & \makecell[l]{$y_t^{\F} = 0.988(11)^{\dagger}$\\$y_t^{\dlm} = 0.9833(32)$} \\

48 & -- & -- & -- & -- & \makecell[l]{$y_t^{\F} = 0.990(20)^{\dagger}$\\$y_t^{\dlm} = 0.9838(67)$} \\

\midrule
&  &  &$\J2 = -0.22$  &  &  \\
\midrule

8 & \makecell[l]{$y_t^{\F} = 0.9328(30)^{\dagger}$\\$y_t^{\dlm} = 0.8781(25)^{\dagger}$} & \makecell[l]{$y_t^{\F} = 0.9356(23)^{\dagger}$\\$y_t^{\dlm} = 0.8907(19)^{\dagger}$} & \makecell[l]{$y_t^{\F} = 0.9364(21)^{\dagger}$\\$y_t^{\dlm} = 0.9019(15)^{\dagger}$} & \makecell[l]{$y_t^{\F} = 0.9410(18)^{\dagger}$\\$y_t^{\dlm} = 0.9112(13)^{\dagger}$} & \makecell[l]{$y_t^{\F} = 0.9461(15)^{\dagger}$\\$y_t^{\dlm} = 0.9167(12)^{\dagger}$} \\

16 & -- & \makecell[l]{$y_t^{\F} = 0.9525(76)$\\$y_t^{\dlm} = 0.9302(42)$} & \makecell[l]{$y_t^{\F} = 0.9513(63)$\\$y_t^{\dlm} = 0.9365(28)$} & \makecell[l]{$y_t^{\F} = 0.9588(45)$\\$y_t^{\dlm} = 0.9431(22)^{\dagger}$} & \makecell[l]{$y_t^{\F} = 0.9644(32)$\\$y_t^{\dlm} = 0.9484(19)^{\dagger}$} \\

24 & -- & -- & \makecell[l]{$y_t^{\F} = 0.948(13)$\\$y_t^{\dlm} = 0.9437(54)$} & \makecell[l]{$y_t^{\F} = 0.9634(71)$\\$y_t^{\dlm} = 0.9525(36)^{\dagger}$} & \makecell[l]{$y_t^{\F} = 0.9697(46)$\\$y_t^{\dlm} = 0.9600(30)^{\dagger}$} \\

32 & -- & -- & -- & \makecell[l]{$y_t^{\F} = 0.971(11)$\\$y_t^{\dlm} = 0.9612(53)$} & \makecell[l]{$y_t^{\F} = 0.9766(68)$\\$y_t^{\dlm} = 0.9702(42)^{\dagger}$} \\

48 & -- & -- & -- & -- & \makecell[l]{$y_t^{\F} = 0.995(16)$\\$y_t^{\dlm} = 0.9905(84)$} \\

\bottomrule
\end{tabularx}

\noindent{\footnotesize{$^\dagger$ $Q$ value below $0.1$.}}

	\end{adjustwidth}
\end{table}

For $\J2 = -0.22$ the estimates for $y_t$ are even further below the Onsager value of $y_t=1$, and again increase with $L_{\min}$ and $L_{\max}$, suggestive of the presence of strong corrections to scaling. Only on the range $[48,88]$, is the result within error bars of the Onsager value. Also here the effectively changing exponent is reflected in poor fit qualities. 
Differently from the previous cases, however, the value for $y_t$ from the Fisher zeros is consistently closer to the expected value than the value from regular FSS, suggesting that corrections to scaling are weaker for the location of the Fisher zeros in this case. The overall worse fit quality for $\dlm$ is also in agreement with stronger correction terms.

\subsubsection{Scaling of Lee-Yang zeros}
In the following we obtain $y_h$ from the scaling behavior (see Equation~\eqref{eq:scalingLY}) of the leading Lee-Yang zero at the (fixed) inverse temperature $\beta_c$, as well as 
from the value of the magnetic susceptibility $\chi$ at $\beta_c$ which follows the scaling relation 
\begin{equation}
	\chi_{L}(\beta_c) = L^{-D} \beta_c \langle M^2 \rangle_{\beta_c} \propto L^{-D+2y_h} = L^{\gamma/\nu},
\end{equation}
with $D$ being the spatial dimension.
In principle, one may also use the pseudo-critical points of the magnetic susceptibility with subtraction and its peak heights which follow the same scaling relation. However, as we use the critical temperature for the Lee-Yang zeros, this would result in an unjust comparison and the values from the magnetic susceptibility by design would be subject to stronger corrections to scaling.

We consider the Lee-Yang zeros at our estimate for the inverse critical temperature $\beta_c$ obtained from the scaling of the real part of the Fisher zeros for the different values of $\J2$.
This estimate is found by applying the fit ansatz $\Re(\beta_0(L)) = \beta_c - a L^{-y_t} - b L^{-2y_t}$, allowing for a first-order correction term and assuming $y_t = 1$.
Using the full range of system sizes, i.e., $L=8,\dots,88$, we obtain $\beta_c J_1 = 1.02350(32), 2.65740(54),  3.25721(42), 4.27017 (59)$ for $\J2=-0.1$, $-0.2$, $-0.21$, and $-0.22$, respectively (cf.\ Supplementary Tables~S9 to S12).
The Lee-Yang zeros $\Im(h_0)$ and the susceptibility $\chi_L$ together with their statistical errors are then evaluated at this value for $\beta_c$.\footnote{Note that as the critical temperature was calculated a posteriori, we have no data for the cumulants at $\beta_c$. In order to estimate the value of the cumulants at $\beta_c$ we use Lagrange interpolation between the four inverse temperatures closest to $\beta_c$, potentially resulting in a small systematic error. As all realizations use the same temperature set the effect of this is not accounted for in the quoted error bars.}
Next, one carries out the FSS analysis on the calculated values for the Lee-Yang zeros and the susceptibility at the obtained $\beta_c$. When doing so the statistical errors in $\Im(h_0)$ and $\chi_L$ are correctly reflected in the statistical error of the critical exponent $y_h$, whereas the uncertainty in $\beta_c$ is not accounted for.
To estimate the influence of the error of $\beta_c$ on the estimate for $y_h$, one may repeat this analysis at $\beta_c-\epsilon(\beta_c)$ and $\beta_c+\epsilon(\beta_c)$ with $\epsilon(\beta_c)$ corresponding to the quoted error bars. This yields estimates $y_h^{(-)}$ and and $y_h^{(+)}$, thus giving rise to a second contribution $|y_h^{(+)}-y_h^{(-)}|/2$ to the error of $y_h$.
Alternatively, to combine both error contributions in the quoted error bars, we carry out jackknifing~\cite{efron1982} over the whole FSS procedure. Specifically, we use the ten jackknife blocks (each containing nine out of ten PA simulation runs), for which we then perform the entire analysis resulting in slightly varying values for the Fisher zeros and the inverse critical temperatures.
The Lee-Yang zeros and the magnetic susceptibilities are analyzed at the respective $\beta_c$ of each jackknife block, and finally result in different estimates for the exponent $y_h$. The final estimates of $y_h$ quoted in Table~\ref{tab:fss_yh} and Supplementary Tables S13 to S20 are then the plain averages of the jackknife blocks.%
Their standard errors are calculated via the usual rescaled variance of the jackknife blocks which accounts for their trivial correlation~\cite{efron1982,Janke2012}. For further details including results for all fitting parameters, see Sections 2.2 and 2.3 of the Supplementary Material. We have checked that the jackknife estimates for $\beta_c J_1$ ($\beta_c J_1=\bcA$, $\bcB$, $\bcC$, and $\bcD$) are in very good agreement with our aforementioned final values.

The expected value for $y_h$ is the $D=2$ Ising value $y_h=1.875$. Table~\ref{tab:fss_yh} summarizes the FSS results using the leading Lee-Yang zero and the magnetic susceptibility for $\J2 = -0.1$, $-0.2$, and $-0.22$.
Also here the results are affected by corrections to scaling reflected in effectively varying exponents that approach the expected value as $L_{\min}$ and $L_{\max}$ increase.
Both methods yield values for $y_h$ well compatible with the expected exponent of $1.875$.

For $\J2=-0.1$ the estimate $y_h^{\LY}$ for $y_h$ from the scaling of the Lee-Yang zeros even on the smallest range of system sizes, i.e., $L\in\{ 8, 16, 24 \}$, is within two error bars of $1.875$, as opposed to the estimate $y_h^\chi$ from the magnetic susceptibility, which is far outside the error margin. 
This indicates stronger corrections to scaling for the magnetic susceptibility, which is also reflected in the overall poorer quality-of-fit value $Q$.
Despite the big difference when including the value for $L=8$, for $L_{\min}>8$ the value for $y_h$ from the Lee-Yang zeros is only marginally closer to the Ising value as compared to the results from the scaling of the magnetic susceptibility.

\begin{table}[H]
    \caption{Jackknife estimates at $\beta_c$ (see text) for $y_h$ from FSS fits using the leading Lee-Yang zero as well as from the value of the magnetic susceptibility $\chi_L$ at $\beta_c$ for different $L_{\min}$ and $L_{\max}$, and using $\J2 = -0.1$, $-0.2$, and $-0.22$.}\label{tab:fss_yh}
	\begin{adjustwidth}{-\extralength}{0cm}
    \begin{tabularx}{\fulllength}{llllll}

\toprule

$\mathbf{L_{\max}}$ & 24 & 32 & 48 & 64 & 88 \\

$\mathbf{L_{\min}}$ &  &  &  &  &  \\

\midrule
&  &  &$\J2 = -0.1$  &  &  \\
\midrule

8 & \makecell[l]{$y_h^{\LY} = 1.87629(97)$\\$y_h^{\chi} = 1.87813(86)$} & \makecell[l]{$y_h^{\LY} = 1.87564(69)$\\$y_h^{\chi} = 1.87730(57)^{\dagger}$} & \makecell[l]{$y_h^{\LY} = 1.8756(13)$\\$y_h^{\chi} = 1.8769(11)^{\dagger}$} & \makecell[l]{$y_h^{\LY} = 1.8754(20)$\\$y_h^{\chi} = 1.8764(17)^{\dagger}$} & \makecell[l]{$y_h^{\LY} = 1.8754(25)$\\$y_h^{\chi} = 1.8763(21)^{\dagger}$} \\

16 & -- & \makecell[l]{$y_h^{\LY} = 1.8747(13)$\\$y_h^{\chi} = 1.8755(11)^{\dagger}$} & \makecell[l]{$y_h^{\LY} = 1.8751(23)$\\$y_h^{\chi} = 1.8757(20)$} & \makecell[l]{$y_h^{\LY} = 1.8750(30)$\\$y_h^{\chi} = 1.8754(26)$} & \makecell[l]{$y_h^{\LY} = 1.8751(35)$\\$y_h^{\chi} = 1.8755(31)$} \\

24 & -- & -- & \makecell[l]{$y_h^{\LY} = 1.8745(38)$\\$y_h^{\chi} = 1.8751(34)$} & \makecell[l]{$y_h^{\LY} = 1.8747(42)$\\$y_h^{\chi} = 1.8750(37)$} & \makecell[l]{$y_h^{\LY} = 1.8750(46)$\\$y_h^{\chi} = 1.8753(40)$} \\

32 & -- & -- & -- & \makecell[l]{$y_h^{\LY} = 1.8758(49)$\\$y_h^{\chi} = 1.8759(42)$} & \makecell[l]{$y_h^{\LY} = 1.8757(50)$\\$y_h^{\chi} = 1.8759(43)$} \\

48 & -- & -- & -- & -- & \makecell[l]{$y_h^{\LY} = 1.8752(49)$\\$y_h^{\chi} = 1.8754(43)$} \\

\midrule
&  &  &$\J2 = -0.2$  &  &  \\
\midrule

8 & \makecell[l]{$y_h^{\LY} = 1.8780(13)$\\$y_h^{\chi} = 1.8828(13)^{\dagger}$} & \makecell[l]{$y_h^{\LY} = 1.8772(16)$\\$y_h^{\chi} = 1.8813(14)^{\dagger}$} & \makecell[l]{$y_h^{\LY} = 1.8771(18)$\\$y_h^{\chi} = 1.8805(16)^{\dagger}$} & \makecell[l]{$y_h^{\LY} = 1.8769(20)$\\$y_h^{\chi} = 1.8801(17)^{\dagger}$} & \makecell[l]{$y_h^{\LY} = 1.8767(30)^{\dagger}$\\$y_h^{\chi} = 1.8789(28)^{\dagger}$} \\

16 & -- & \makecell[l]{$y_h^{\LY} = 1.8750(26)$\\$y_h^{\chi} = 1.8767(23)$} & \makecell[l]{$y_h^{\LY} = 1.8757(28)$\\$y_h^{\chi} = 1.8767(24)$} & \makecell[l]{$y_h^{\LY} = 1.8757(32)$\\$y_h^{\chi} = 1.8765(28)$} & \makecell[l]{$y_h^{\LY} = 1.8759(44)$\\$y_h^{\chi} = 1.8763(38)$} \\

24 & -- & -- & \makecell[l]{$y_h^{\LY} = 1.8759(35)$\\$y_h^{\chi} = 1.8761(30)$} & \makecell[l]{$y_h^{\LY} = 1.8757(38)$\\$y_h^{\chi} = 1.8759(33)$} & \makecell[l]{$y_h^{\LY} = 1.8760(48)$\\$y_h^{\chi} = 1.8761(42)$} \\

32 & -- & -- & -- & \makecell[l]{$y_h^{\LY} = 1.8765(41)$\\$y_h^{\chi} = 1.8765(36)$} & \makecell[l]{$y_h^{\LY} = 1.8763(52)$\\$y_h^{\chi} = 1.8763(46)$} \\

48 & -- & -- & -- & -- & \makecell[l]{$y_h^{\LY} = 1.8758(66)$\\$y_h^{\chi} = 1.8759(57)$} \\

\midrule
&  &  &$\J2 = -0.22$  &  &  \\
\midrule

8 & \makecell[l]{$y_h^{\LY} = 1.88294(89)$\\$y_h^{\chi} = 1.88778(65)^{\dagger}$} & \makecell[l]{$y_h^{\LY} = 1.8818(11)^{\dagger}$\\$y_h^{\chi} = 1.8860(11)^{\dagger}$} & \makecell[l]{$y_h^{\LY} = 1.8808(12)^{\dagger}$\\$y_h^{\chi} = 1.8845(12)^{\dagger}$} & \makecell[l]{$y_h^{\LY} = 1.8804(10)^{\dagger}$\\$y_h^{\chi} = 1.88300(95)^{\dagger}$} & \makecell[l]{$y_h^{\LY} = 1.8803(10)^{\dagger}$\\$y_h^{\chi} = 1.88288(95)^{\dagger}$} \\

16 & -- & \makecell[l]{$y_h^{\LY} = 1.8777(22)$\\$y_h^{\chi} = 1.8791(19)$} & \makecell[l]{$y_h^{\LY} = 1.8766(16)$\\$y_h^{\chi} = 1.8777(16)$} & \makecell[l]{$y_h^{\LY} = 1.8781(14)$\\$y_h^{\chi} = 1.8786(13)$} & \makecell[l]{$y_h^{\LY} = 1.8781(13)$\\$y_h^{\chi} = 1.8786(12)$} \\

24 & -- & -- & \makecell[l]{$y_h^{\LY} = 1.8752(17)$\\$y_h^{\chi} = 1.8758(16)$} & \makecell[l]{$y_h^{\LY} = 1.8779(18)$\\$y_h^{\chi} = 1.8781(15)$} & \makecell[l]{$y_h^{\LY} = 1.8780(17)$\\$y_h^{\chi} = 1.8782(15)$} \\

32 & -- & -- & -- & \makecell[l]{$y_h^{\LY} = 1.8792(23)$\\$y_h^{\chi} = 1.8792(19)$} & \makecell[l]{$y_h^{\LY} = 1.8793(21)$\\$y_h^{\chi} = 1.8792(18)$} \\

48 & -- & -- & -- & -- & \makecell[l]{$y_h^{\LY} = 1.8827(36)$\\$y_h^{\chi} = 1.8817(30)$} \\

\bottomrule
\end{tabularx}

\noindent{\footnotesize{$^\dagger$ $Q$ value below $0.1$.}}

	\end{adjustwidth}
\end{table}

Also for $\J2=-0.2$ the difference between the methods shows most clearly when including the value for $L=8$. 
Here the estimate from the partition function zeros is much closer to the expected one, albeit still not within error bars.
When choosing $L_{\min} = 16$, the value from the Lee-Yang zeros is always within at most two error bars of the expected value, which is not the case for the value from the magnetic susceptibility. However, this observation may not be significant as both values are within error bars.
For $L \geq 24$, both methods yield values compatible with each other and with $1.875$.
Similarly, for $\J2 =-0.22$ the Lee-Yang value for $y_h$ is much closer to $1.875$ (but again not within error bars) than the magnetic susceptibility one when including $L=8$. When excluding the smallest system size, both methods yield results compatible with the Ising exponent, and both methods appear to perform equally well.

\section{Conclusions}

We have studied the Fisher and Lee-Yang zeros for the frustrated $J_1$-$J_2$ Ising model on the honeycomb lattice. 
The partition function zeros are obtained using a recently suggested cumulant method~\cite{Flindt2013,Deger2018,Deger2019,Deger2020} that does not require knowledge of the density of states $\Omega$.
For small systems where $\Omega$ was available, we compared the values for the leading Fisher and Lee-Yang zeros from the cumulant method against the directly obtained estimates and observed only small deviations that vanish exponentially in the cumulant order $n$, regardless of the value of $\J2$. For larger systems, we also saw an exponential convergence of the cumulant estimates to their asymptotic values when evaluating the cumulants at $\beta$ close to $\Re(\beta_0)$.

We compared FSS using the location of the leading partition function zeros with a traditional FSS protocol. Both approaches indicate that the model remains in the Ising universality class for all studied values of $\J2 > -1/4$.
For the temperature exponent $y_t$, our numerical results do not favor one method over the other.
Instead, both approaches seem to be subject to non-trivial corrections to scaling, such that depending on $\J2$ one or the other approach appears preferable.
For $\J2=-0.1$ conventional FSS shows practically no signs of corrections to scaling even for very small systems, whereas the values obtained for $y_t$ from the partition function zeros have a clear system-size dependence.
On the other hand, for values of $\J2$ closer to $-1/4$, conventional FSS is subject to very strong corrections to scaling, whereas the values obtained using the partition function zeros show only slightly stronger corrections as compared to $\J2=-0.1$. Therefore, our data for the partition function zeros for $\J2 = -0.22$ convincingly indicate that the system remains in the Ising universality class, whereas the results from traditional FSS alone were much less conclusive. 
The field exponent $y_h$ as obtained from the Lee-Yang zeros in most cases was marginally closer to the expected value than the estimate derived from the magnetic susceptibility, although only when including the smallest system size of $L=8$ the former approach performed significantly better than the latter.

Thus, based on our results, studying the critical behavior using the partition function zeros does not in general promise to yield results less afflicted by scaling corrections but, as expected, in different regimes one or the other approach might have an edge in this respect. Both techniques (as well as other scaling paradigms) can hence be used with good success in a complementary fashion.

\authorcontributions{Conceptualization, D.G. and W.J.; methodology, D.G.; software, D.G.; validation, W.J. and M.W.; formal analysis, D.G.; writing---original draft preparation, D.G.; writing---review and editing, M.W. and W.J.; visualization, D.G.; supervision, M.W. and W.J.; project administration, W.J.; funding acquisition, W.J. All authors have read and agreed to the published version of the manuscript.}

\funding{This project was supported by the Deutsch-Franz\"osische Hochschule (DFH-UFA) through the Doctoral College ``$\mathbb{L}^4$'' under Grant No.\ CDFA-02-07. We further acknowledge the resources provided by the Leipzig Graduate School of Natural Sciences ``BuildMoNa''.}

\dataavailability{The data that support the findings of this study are available from the corresponding author upon reasonable request.} 

\acknowledgments{We thank  Le\"ila Moueddene for interesting discussions. -- This work is dedicated to the memory of Ralph Kenna who passed away at the end of October 2023. Ralph was fascinated by partition function zeros since his PhD Thesis work~\cite{Kenna1991,Kenna1994} and revisited this versatile tool for obtaining properties of phase transitions during his scientific career many times~\cite{Moueddene2024}. We will remember Ralph as good friend and inspiring collaborator. His deep insights, witty comments, and captivating presentations will be dearly missed.}

\conflictsofinterest{The authors declare no conflicts of interest. The funders had no role in the design of the study; in the collection, analyses, or interpretation of data; in the writing of the manuscript; or in the decision to publish the results.} 

\abbreviations{Abbreviations}{
The following abbreviations are used in this manuscript:\\

\noindent 
\begin{tabular}{@{}ll}
MC & Monte Carlo\\
PA & Population annealing\\
LY & Lee-Yang\\
F & Fisher\\
FSS & Finite-size scaling\\
RG & Renormalization group
\end{tabular}

}

\appendixtitles{no} %
\appendixstart
\appendix

\section[\appendixname~\thesection]{Fisher zeros for $\J2=0$}\label{sec:FisherZeros_J2_0}
As a consistency check,  we carried out the partition function zero analysis for $\J2=0$ in the complex temperature plane, where exact Fisher zeros are known in the thermodynamic limit~\cite{Matveev1996,Kim2008b}. The Fisher zeros of the field-free Ising model on the honeycomb lattice without next-nearest-neighbor interactions lie on the partial circle $z^2 = e^{i \varphi}$ for $\varphi \in [\pi/3,5 \pi/3]$, and the heart-shape-like curve given by~\cite{Kim2008b}
\begin{equation}
{\left[\Im(z^2)\right]}^2 = 1 + 2 \Re(z^2) - {\left[\Re(z^2)\right]}^2 \pm \sqrt{8\Re(z^2)}.
\end{equation} 
\begin{figure}[H]
	\includegraphics{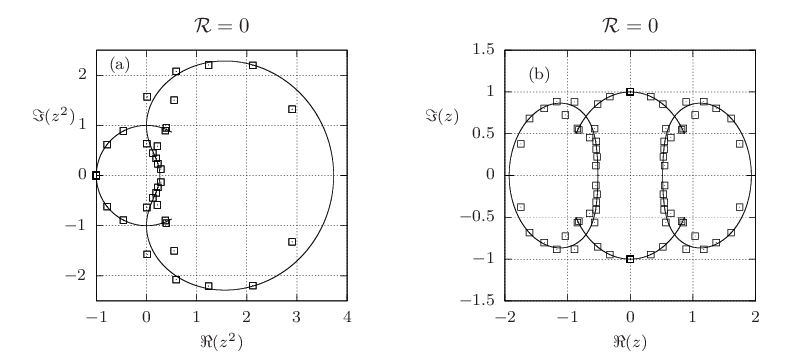}
	\caption{Fisher zeros in the complex temperature planes (a) $z^2 = e^{-2\beta J_1}$ and (b) $z = e^{-\beta J_1}$ for $\J2=0$. Points denote our data for $L=4$ obtained by exact enumeration. The solid lines are the exact zeros for $L\rightarrow\infty$~\cite{Matveev1996,Kim2008b}.\label{fig:allZeros_L4_J2_0}}
\end{figure}
Figure~\ref{fig:allZeros_L4_J2_0} shows the Fisher zeros for $L=4$ (points) and for $L\rightarrow\infty$ (solid lines). Despite the rather small system size, both appear to be already somewhat compatible, with the closest zero for the finite system size, of course, not being located on the positive real axis.
As in the absence of next-nearest-neighbor interactions the partition function is an even polynomial in $z$, the natural variable to consider the Fisher zeros is $z^2$ [see panel~(a)].
When using $z$ as variable, the $z\rightarrow -z$ symmetry is visible [see panel (b)], which was absent for $\J2 \neq 0$, cf.\ Figure~\ref{fig:allZeros_L4}.

\section[\appendixname~\thesection]{Convergence of the cumulant method for larger system sizes}\label{app:convergenceCumulants}
We have tested the convergence of the cumulant method for small system sizes extensively to assure its convergence. For larger system sizes we did not have access to the density of states.
To test the convergence for these larger system sizes, we therefore consider the predicted value of the partition function zero as a function of cumulant order~$n$.

\begin{figure}[H]
	\includegraphics{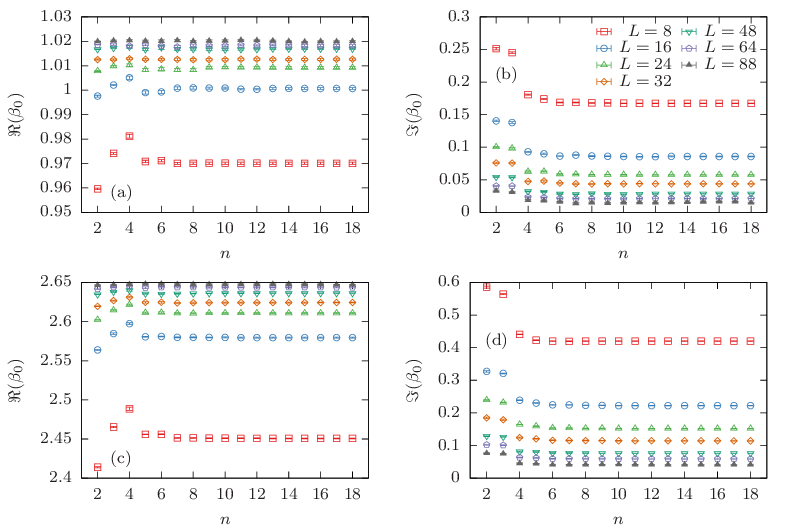}
	\caption{Real and imaginary part of the Fisher zeros $\beta_0$ obtained through the cumulant method as a function of $n$ [see Equation~(\ref{eq:cumulantMethod})] for different $L$. In (a) and (b) $\J2=-0.1$, and in (c) and (d) $\J2=-0.2$.\label{fig:cumulantsDiffJ2_and_L}}
\end{figure}

Figure~\ref{fig:cumulantsDiffJ2_and_L} shows the estimated location of the Fisher zero using the cumulant method. The $\beta_0$ is found by determining the zero crossing of $\Re(\beta-\beta_0)$, and the imaginary part is calculated using the equation for $|\beta-\beta_0|^2$ at that value for $\beta$. 
For all system sizes and for all considered choices of $\J2$, the estimated value from the cumulant method quickly approaches a constant with increasing $n$.

\begin{figure}[H]
	\includegraphics{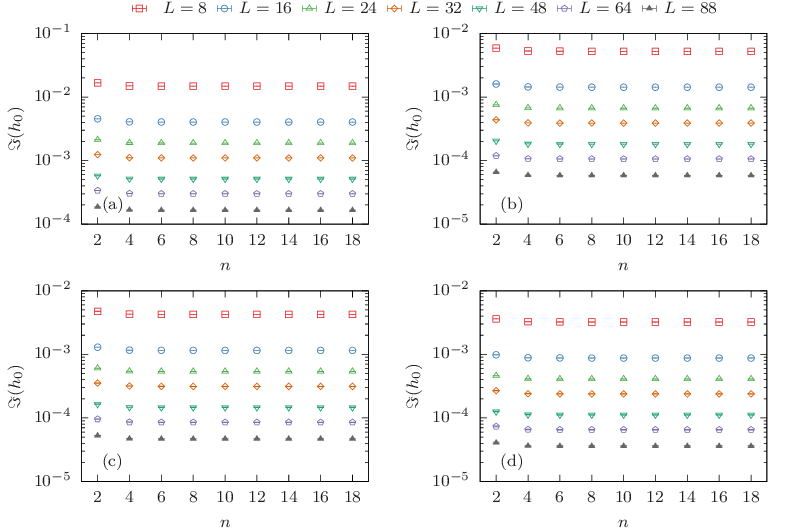}
	\caption{Imaginary part of the Lee-Yang zeros $h_0$ for different system sizes $L$ and coupling strengths $\J2$. (a) $\J2=-0.1$, (b) $\J2=-0.2$, (c) $\J2=-0.21$, and (d) $\J2=-0.22$.\label{fig:cumulantsLYDiffJ2_and_L}}
\end{figure}

Similarly, in Figure~\ref{fig:cumulantsLYDiffJ2_and_L} we show the values for the complex field $h_0$ of the Lee-Yang zeros at $\beta_c$ estimated by the cumulant method for the different values of $\J2$ as a function of $n$. These also converge quickly.
Note that even on the logarithmic scale of Figure~\ref{fig:cumulantsLYDiffJ2_and_L}, error bars do not appear to grow significantly as the order of cumulants is increased.
As mentioned above, we attribute this to the fact that despite increasing statistical errors of the individual cumulants their ratios show only little statistical fluctuation due to the cross-correlation between the terms.

\begin{adjustwidth}{-\extralength}{0cm}

\reftitle{References}

\PublishersNote{}
\end{adjustwidth}


\begin{thebibliography}{999}

\bibitem[Diep(2013)]{Diep2012}
Diep, H.T., Ed.
\newblock {\em Frustrated Spin Systems}, 2nd ed. ed.; World Scientific:
  Singapore,  2013.
\newblock {\url{https://doi.org/10.1142/8676}}.

\bibitem[Azaria et~al.(1987)Azaria, Diep, and Giacomini]{Azaria1987}
Azaria, P.; Diep, H.T.; Giacomini, H.
\newblock Coexistence of order and disorder and reentrance in an exactly
  solvable model.
\newblock {\em Phys. Rev. Lett.} {\bf 1987}, {\em 59},~1629--1632.
\newblock {\url{https://doi.org/10.1103/physrevlett.59.1629}}.

\bibitem[Espriu and Prats(2004)]{Espriu2004}
Espriu, D.; Prats, A.
\newblock Dynamics of the two-dimensional gonihedric spin model.
\newblock {\em Phys. Rev. E} {\bf 2004}, {\em 70},~046117.
\newblock {\url{https://doi.org/10.1103/physreve.70.046117}}.

\bibitem[Mueller et~al.(2014)Mueller, Johnston, and Janke]{Mueller2014}
Mueller, M.; Johnston, D.A.; Janke, W.
\newblock Multicanonical analysis of the plaquette-only gonihedric {I}sing
  model and its dual.
\newblock {\em Nucl. Phys. B} {\bf 2014}, {\em 888},~214--235.
\newblock {\url{https://doi.org/10.1016/j.nuclphysb.2014.09.009}}.

\bibitem[Nightingale(1977)]{nightingale:77}
Nightingale, M.
\newblock Non-universality for {I}sing-like spin systems.
\newblock {\em Phys. Lett. A} {\bf 1977}, {\em 59},~486--488.
\newblock {\url{https://doi.org/10.1016/0375-9601(77)90665-X}}.

\bibitem[Jin et~al.(2012)Jin, Sen, and Sandvik]{Jin2012}
Jin, S.; Sen, A.; Sandvik, A.W.
\newblock Ashkin-{T}eller criticality and pseudo-first-order behavior in a
  frustrated {I}sing model on the square lattice.
\newblock {\em Phys. Rev. Lett.} {\bf 2012}, {\em 108},~045702.
\newblock {\url{https://doi.org/10.1103/physrevlett.108.045702}}.

\bibitem[Kalz et~al.(2011)Kalz, Honecker, and Moliner]{Kalz2011}
Kalz, A.; Honecker, A.; Moliner, M.
\newblock Analysis of the phase transition for the {I}sing model on the
  frustrated square lattice.
\newblock {\em Phys. Rev. B} {\bf 2011}, {\em 84},~174407.
\newblock {\url{https://doi.org/10.1103/physrevb.84.174407}}.

\bibitem[Kalz and Honecker(2012)]{Kalz2012}
Kalz, A.; Honecker, A.
\newblock Location of the Potts-critical end point in the frustrated {I}sing
  model on the square lattice.
\newblock {\em Phys. Rev. B} {\bf 2012}, {\em 86},~134410.
\newblock {\url{https://doi.org/10.1103/physrevb.86.134410}}.

\bibitem[Yoshiyama and Hukushima(2023)]{Yoshiyama2023}
Yoshiyama, K.; Hukushima, K.
\newblock Higher-order tensor renormalization group study of the
  ${J_1}$-${J_2}$ {I}sing model on a square lattice.
\newblock {\em Phys. Rev. E} {\bf 2023}, {\em 108},~054124.
\newblock {\url{https://doi.org/10.1103/physreve.108.054124}}.

\bibitem[Bob{\'{a}}k et~al.(2016)Bob{\'{a}}k, Lu{\v{c}}ivjansk{\'{y}},
  {\v{Z}}ukovi{\v{c}}, Borovsk{\'{y}}, and Balcerzak]{Bobak2016}
Bob{\'{a}}k, A.; Lu{\v{c}}ivjansk{\'{y}}, T.; {\v{Z}}ukovi{\v{c}}, M.;
  Borovsk{\'{y}}, M.; Balcerzak, T.
\newblock Tricritical behaviour of the frustrated {I}sing antiferromagnet on
  the honeycomb lattice.
\newblock {\em Phys. Lett. A} {\bf 2016}, {\em 380},~2693--2697.
\newblock {\url{https://doi.org/10.1016/j.physleta.2016.06.019}}.

\bibitem[Acevedo et~al.(2021)Acevedo, Arlego, and Lamas]{Acevedo2021}
Acevedo, S.; Arlego, M.; Lamas, C.A.
\newblock Phase diagram study of a two-dimensional frustrated antiferromagnet
  via unsupervised machine learning.
\newblock {\em Phys. Rev. B} {\bf 2021}, {\em 103},~134422.
\newblock {\url{https://doi.org/10.1103/physrevb.103.134422}}.

\bibitem[{\v{Z}}ukovi{\v{c}}(2021)]{Zukovic2021}
{\v{Z}}ukovi{\v{c}}, M.
\newblock Critical properties of the frustrated {I}sing model on a honeycomb
  lattice: A {M}onte {C}arlo study.
\newblock {\em Phys. Lett. A} {\bf 2021}, {\em 404},~127405.
\newblock {\url{https://doi.org/10.1016/j.physleta.2021.127405}}.

\bibitem[Schmidt and Godoy(2021)]{Schmidt2021}
Schmidt, M.; Godoy, P.
\newblock Phase transitions in the Ising antiferromagnet on the frustrated
  honeycomb lattice.
\newblock {\em J. Magn. Magn. Mater.} {\bf 2021}, {\em 537},~168151.
\newblock {\url{https://doi.org/10.1016/j.jmmm.2021.168151}}.

\bibitem[Janke(2007)]{Janke2008}
Janke, W., Ed.
\newblock {\em Rugged Free Energy Landscapes -- Common Computational Approaches
  to Spin Glasses, Structural Glasses and Biological Macromolecules}; Vol. 736,
  {\em Lecture Notes in Physics}, Springer: Berlin,  2007.
\newblock {\url{https://doi.org/10.1007/978-3-540-74029-2}}.

\bibitem[Iba(2001)]{Iba2001}
Iba, Y.
\newblock Population {M}onte {C}arlo algorithms.
\newblock {\em Transactions of the Japanese Society for Artificial
  Intelligence} {\bf 2001}, {\em 16},~279--286.
\newblock {\url{https://doi.org/10.1527/tjsai.16.279}}.

\bibitem[Hukushima and Iba(2003)]{Hukushima2003}
Hukushima, K.; Iba, Y.
\newblock Population annealing and its application to a spin glass.
\newblock {\em {AIP} Conf. Proc.} {\bf 2003}, {\em 690},~200--206.
\newblock {\url{https://doi.org/10.1063/1.1632130}}.

\bibitem[Machta(2010)]{Machta2010}
Machta, J.
\newblock Population annealing with weighted averages: A {M}onte {C}arlo method
  for rough free-energy landscapes.
\newblock {\em Phys. Rev. E} {\bf 2010}, {\em 82},~026704.
\newblock {\url{https://doi.org/10.1103/physreve.82.026704}}.

\bibitem[Wang et~al.(2015)Wang, Machta, and Katzgraber]{Wang2015}
Wang, W.; Machta, J.; Katzgraber, H.G.
\newblock Population annealing: Theory and application in spin glasses.
\newblock {\em Phys. Rev. E} {\bf 2015}, {\em 92}.
\newblock {\url{https://doi.org/10.1103/physreve.92.063307}}.

\bibitem[Christiansen et~al.(2019)Christiansen, Weigel, and
  Janke]{Christiansen2019a}
Christiansen, H.; Weigel, M.; Janke, W.
\newblock Accelerating molecular dynamics simulations with population
  annealing.
\newblock {\em Phys. Rev. Lett.} {\bf 2019}, {\em 122},~060602.
\newblock {\url{https://doi.org/10.1103/physrevlett.122.060602}}.

\bibitem[Rose and Machta(2019)]{Rose2019}
Rose, N.; Machta, J.
\newblock Equilibrium microcanonical annealing for first-order phase
  transitions.
\newblock {\em Phys. Rev. E} {\bf 2019}, {\em 100},~063304.
\newblock {\url{https://doi.org/10.1103/physreve.100.063304}}.

\bibitem[Yang and Lee(1952)]{Yang1952}
Yang, C.N.; Lee, T.D.
\newblock Statistical theory of equations of state and phase transitions. I.
  Theory of condensation.
\newblock {\em Phys. Rev.} {\bf 1952}, {\em 87},~404--409.
\newblock {\url{https://doi.org/10.1103/physrev.87.404}}.

\bibitem[Lee and Yang(1952)]{Lee1952}
Lee, T.D.; Yang, C.N.
\newblock Statistical theory of equations of state and phase transitions. II.
  Lattice gas and Ising model.
\newblock {\em Phys. Rev.} {\bf 1952}, {\em 87},~410--419.
\newblock {\url{https://doi.org/10.1103/physrev.87.410}}.

\bibitem[Fisher(1965)]{fisher1965nature}
Fisher, M.E.
\newblock The nature of critical points. In {\em Lecture Notes in Theoretical
  Physics}; University of Colorado Press: Boulder,  1965; Vol. VII C.

\bibitem[Janke and Kenna(2001)]{Janke2001}
Janke, W.; Kenna, R.
\newblock The strength of first and second order phase transitions from
  partition function zeroes.
\newblock {\em J. Stat. Phys.} {\bf 2001}, {\em 102},~1211--1227.
\newblock {\url{https://doi.org/10.1023/a:1004836227767}}.

\bibitem[Janke and Kenna(2002{\natexlab{a}})]{Janke2002b}
Janke, W.; Kenna, R., Analysis of the density of partition function zeroes: A
  measure for phase transition strength.
\newblock In {\em Computer Simulation Studies in Condensed-Matter Physics XIV};
  Landau, D.P.; Lewis, S.P.; Schüttler, H.B., Eds.; Springer: Berlin
  Heidelberg,  2002; pp. 97--101.
\newblock {\url{https://doi.org/10.1007/978-3-642-59406-9_14}}.

\bibitem[Janke and Kenna(2002{\natexlab{b}})]{Janke2002a}
Janke, W.; Kenna, R.
\newblock Density of partition function zeroes and phase transition strength.
\newblock {\em Comput. Phys. Commun.} {\bf 2002}, {\em 147},~443--446.
\newblock {\url{https://doi.org/10.1016/s0010-4655(02)00323-5}}.

\bibitem[Janke and Kenna(2002{\natexlab{c}})]{Janke2002}
Janke, W.; Kenna, R.
\newblock Phase transition strengths from the density of partition function
  zeroes.
\newblock {\em Nucl. Phys. B Proc. Suppl.} {\bf 2002}, {\em
  106–107},~905--907.
\newblock {\url{https://doi.org/10.1016/s0920-5632(01)01881-3}}.

\bibitem[Janke and Kenna(2002{\natexlab{d}})]{Janke2002c}
Janke, W.; Kenna, R.
\newblock Finite-size scaling and corrections in the Ising model with
  Brascamp-Kunz boundary conditions.
\newblock {\em Phys. Rev. B} {\bf 2002}, {\em 65},~064110.
\newblock {\url{https://doi.org/10.1103/physrevb.65.064110}}.

\bibitem[Janke and Kenna(2002{\natexlab{e}})]{Janke2002d}
Janke, W.; Kenna, R.
\newblock Exact finite-size scaling with corrections in the two-dimensional
  Ising model with special boundary conditions.
\newblock {\em Nucl. Phys. B Proc. Suppl.} {\bf 2002}, {\em
  106–107},~929--931.
\newblock {\url{https://doi.org/10.1016/s0920-5632(01)01889-8}}.

\bibitem[Janke et~al.(2003)Janke, Johnston, and Kenna]{Janke2003}
Janke, W.; Johnston, D.A.; Kenna, R.
\newblock New methods to measure phase transition strength.
\newblock {\em Nucl. Phys. B Proc. Suppl.} {\bf 2003}, {\em 119},~882--884.
\newblock {\url{https://doi.org/10.1016/s0920-5632(03)01710-9}}.

\bibitem[Janke et~al.(2004)Janke, Johnston, and Kenna]{Janke2004}
Janke, W.; Johnston, D.A.; Kenna, R.
\newblock Phase transition strength through densities of general distributions
  of zeroes.
\newblock {\em Nucl. Phys. B} {\bf 2004}, {\em 682},~618--634.
\newblock {\url{https://doi.org/10.1016/j.nuclphysb.2004.01.028}}.

\bibitem[Janke et~al.(2005{\natexlab{a}})Janke, Johnston, and Kenna]{Janke2005}
Janke, W.; Johnston, D.A.; Kenna, R.
\newblock Critical exponents from general distributions of zeroes.
\newblock {\em Comput. Phys. Commun.} {\bf 2005}, {\em 169},~457--461.
\newblock {\url{https://doi.org/10.1016/j.cpc.2005.03.101}}.

\bibitem[Janke et~al.(2005{\natexlab{b}})Janke, Johnston, and Kenna]{Kenna2005}
Janke, W.; Johnston, D.A.; Kenna, R.
\newblock Properties of phase transitions of higher-order.
\newblock {\em PoS (LAT2005)} {\bf 2005}, {\em 244}.
\newblock {\url{https://doi.org/10.22323/1.020.0244}}.

\bibitem[Janke et~al.(2006)Janke, Johnston, and Kenna]{Janke2006}
Janke, W.; Johnston, D.A.; Kenna, R.
\newblock Properties of higher-order phase transitions.
\newblock {\em Nucl. Phys. B} {\bf 2006}, {\em 736},~319--328.
\newblock {\url{https://doi.org/10.1016/j.nuclphysb.2005.12.013}}.

\bibitem[Kenna et~al.(2006{\natexlab{a}})Kenna, Johnston, and Janke]{Kenna2006}
Kenna, R.; Johnston, D.A.; Janke, W.
\newblock Scaling relations for logarithmic corrections.
\newblock {\em Phys. Rev. Lett.} {\bf 2006}, {\em 96},~115701.
\newblock {\url{https://doi.org/10.1103/physrevlett.96.115701}}.

\bibitem[Kenna et~al.(2006{\natexlab{b}})Kenna, Johnston, and
  Janke]{Kenna2006a}
Kenna, R.; Johnston, D.A.; Janke, W.
\newblock Self-consistent scaling theory for logarithmic-correction exponents.
\newblock {\em Phys. Rev. Lett.} {\bf 2006}, {\em 97},~155702.
\newblock {\url{https://doi.org/10.1103/physrevlett.97.155702}}.

\bibitem[Moueddene et~al.(2024)Moueddene, Donoso, and Berche]{Moueddene2024a}
Moueddene, L.; Donoso, A.; Berche, B.
\newblock Ralph {K}enna's scaling relations in critical phenomena.
\newblock {\em Entropy} {\bf 2024}, {\em 26},~221.
\newblock {\url{https://doi.org/10.3390/e26030221}}.

\bibitem[Deger and Flindt(2019)]{Deger2019}
Deger, A.; Flindt, C.
\newblock Determination of universal critical exponents using Lee-Yang theory.
\newblock {\em Phys. Rev. Research} {\bf 2019}, {\em 1},~023004.
\newblock {\url{https://doi.org/10.1103/physrevresearch.1.023004}}.

\bibitem[Moueddene et~al.(2024)Moueddene, G~Fytas, Holovatch, Kenna, and
  Berche]{Moueddene2024}
Moueddene, L.; G~Fytas, N.; Holovatch, Y.; Kenna, R.; Berche, B.
\newblock Critical and tricritical singularities from small-scale {M}onte
  {C}arlo simulations: the {B}lume-{C}apel model in two dimensions.
\newblock {\em J. Stat. Mech: Theory Exp.} {\bf 2024}, {\em 2024},~023206.
\newblock {\url{https://doi.org/10.1088/1742-5468/ad1d60}}.

\bibitem[Monroe and Kim(2007)]{Monroe2007}
Monroe, J.L.; Kim, S.Y.
\newblock Phase diagram and critical exponent $\nu$ for the nearest-neighbor
  and next-nearest-neighbor interaction {I}sing model.
\newblock {\em Phys. Rev. E} {\bf 2007}, {\em 76},~021123.
\newblock {\url{https://doi.org/10.1103/physreve.76.021123}}.

\bibitem[Kim(2015)]{Kim2015}
Kim, S.Y.
\newblock Ising antiferromagnet on a finite triangular lattice with free
  boundary conditions.
\newblock {\em J. Korean Phys. Soc.} {\bf 2015}, {\em 67},~1517--1523.
\newblock {\url{https://doi.org/10.3938/jkps.67.1517}}.

\bibitem[Sarkanych et~al.(2015)Sarkanych, Holovatch, and Kenna]{Sarkanych2015}
Sarkanych, P.; Holovatch, Y.; Kenna, R.
\newblock On the phase diagram of the 2d {I}sing model with frustrating dipole
  interaction.
\newblock {\em Ukrainian J. Phys.} {\bf 2015}, {\em 60},~334--338.
\newblock {\url{https://doi.org/10.15407/ujpe60.04.0334}}.

\bibitem[Kim(2021)]{Kim2021}
Kim, S.Y.
\newblock Study of the frustrated Ising model on a square lattice based on the
  exact density of states.
\newblock {\em J. Korean Phys. Soc.} {\bf 2021}, {\em 79},~894--902.
\newblock {\url{https://doi.org/10.1007/s40042-021-00296-8}}.

\bibitem[Peng et~al.(2015)Peng, Zhou, Wei, Cui, Du, and Liu]{Peng2015}
Peng, X.; Zhou, H.; Wei, B.B.; Cui, J.; Du, J.; Liu, R.B.
\newblock Experimental observation of Lee-Yang zeros.
\newblock {\em Phys. Rev. Lett.} {\bf 2015}, {\em 114},~010601.
\newblock {\url{https://doi.org/10.1103/physrevlett.114.010601}}.

\bibitem[Ananikian and Kenna(2015)]{Ananikian2015}
Ananikian, N.; Kenna, R.
\newblock Imaginary magnetic fields in the real world.
\newblock {\em Physics} {\bf 2015}, {\em 8},~2.
\newblock {\url{https://doi.org/10.1103/physics.8.2}}.

\bibitem[Fl{\"a}schner et~al.(2018)Fl{\"a}schner, Vogel, Tarnowski, Rem,
  L{\"u}hmann, Heyl, Budich, Mathey, Sengstock, and Weitenberg]{flaschner:18}
Fl{\"a}schner, N.; Vogel, D.; Tarnowski, M.; Rem, B.; L{\"u}hmann, D.S.; Heyl,
  M.; Budich, J.; Mathey, L.; Sengstock, K.; Weitenberg, C.
\newblock Observation of dynamical vortices after quenches in a system with
  topology.
\newblock {\em Nat. Phys.} {\bf 2018}, {\em 14},~265--268.
\newblock {\url{https://doi.org/10.1038/s41567-017-0013-8}}.

\bibitem[Mac\^edo et~al.(2023)Mac\^edo, Vasilopoulos, Akritidis, Plascak,
  Fytas, and Weigel]{macedo:23}
Mac\^edo, A.R.S.; Vasilopoulos, A.; Akritidis, M.; Plascak, J.A.; Fytas, N.G.;
  Weigel, M.
\newblock Two-dimensional dilute Baxter-Wu model: Transition order and
  universality.
\newblock {\em Phys. Rev. E} {\bf 2023}, {\em 108},~024140.
\newblock {\url{https://doi.org/10.1103/PhysRevE.108.024140}}.

\bibitem[Flindt and Garrahan(2013)]{Flindt2013}
Flindt, C.; Garrahan, J.P.
\newblock Trajectory phase transitions, Lee-Yang zeros, and high-order
  cumulants in full counting statistics.
\newblock {\em Phys. Rev. Lett.} {\bf 2013}, {\em 110},~050601.
\newblock {\url{https://doi.org/10.1103/physrevlett.110.050601}}.

\bibitem[Itzykson et~al.(1983)Itzykson, Pearson, and Zuber]{Itzykson1983}
Itzykson, C.; Pearson, R.B.; Zuber, J.B.
\newblock Distribution of zeros in {I}sing and gauge models.
\newblock {\em Nucl. Phys. B} {\bf 1983}, {\em 220},~415--433.
\newblock {\url{https://doi.org/10.1016/0550-3213(83)90499-6}}.

\bibitem[Bena et~al.(2005)Bena, Droz, and Lipowski]{Bena2005}
Bena, I.; Droz, M.; Lipowski, A.
\newblock Statistical mechanics of equilibrium and nonequilibrium phase
  transitions: The {Y}ang–{L}ee formalism.
\newblock {\em Int. J. Mod. Phys. B} {\bf 2005}, {\em 19},~4269--4329.
\newblock {\url{https://doi.org/10.1142/s0217979205032759}}.

\bibitem[Fröhlich and Rodriguez(2012)]{Froehlich2012}
Fröhlich, J.; Rodriguez, P.F.
\newblock Some applications of the Lee-Yang theorem.
\newblock {\em J. Math. Phys.} {\bf 2012}, {\em 53}.
\newblock {\url{https://doi.org/10.1063/1.4749391}}.

\bibitem[Krasnytska et~al.(2015)Krasnytska, Berche, Holovatch, and
  Kenna]{Krasnytska2015}
Krasnytska, M.; Berche, B.; Holovatch, Y.; Kenna, R.
\newblock Violation of {L}ee-{Y}ang circle theorem for {I}sing phase
  transitions on complex networks.
\newblock {\em Europhys. Lett.} {\bf 2015}, {\em 111},~60009.
\newblock {\url{https://doi.org/10.1209/0295-5075/111/60009}}.

\bibitem[Katsura et~al.(1971)Katsura, Abe, and Yamamoto]{Katsura1971}
Katsura, S.; Abe, Y.; Yamamoto, M.
\newblock Distribution of zeros of the partition function of the {I}sing model.
\newblock {\em J. Phys. Soc. Japan} {\bf 1971}, {\em 30},~347--357.
\newblock {\url{https://doi.org/10.1143/jpsj.30.347}}.

\bibitem[Kenna and Lang(1991)]{Kenna1991}
Kenna, R.; Lang, C.
\newblock Finite size scaling and the zeroes of the partition function in the
  $\Phi_4^4$ model.
\newblock {\em Phys. Lett. B} {\bf 1991}, {\em 264},~396--400.
\newblock {\url{https://doi.org/10.1016/0370-2693(91)90367-y}}.

\bibitem[Kenna and Lang(1994)]{Kenna1994}
Kenna, R.; Lang, C.B.
\newblock Scaling and density of {L}ee-{Y}ang zeros in the four-dimensional
  {I}sing model.
\newblock {\em Phys. Rev. E} {\bf 1994}, {\em 49},~5012--5017.
\newblock {\url{https://doi.org/10.1103/physreve.49.5012}}.

\bibitem[Hong and Kim(2020)]{Hong2020}
Hong, S.; Kim, D.H.
\newblock Logarithmic finite-size scaling correction to the leading Fisher
  zeros in the $p$-state clock model: A higher-order tensor renormalization
  group study.
\newblock {\em Phys. Rev. E} {\bf 2020}, {\em 101},~012124.
\newblock {\url{https://doi.org/10.1103/physreve.101.012124}}.

\bibitem[Deger et~al.(2018)Deger, Brandner, and Flindt]{Deger2018}
Deger, A.; Brandner, K.; Flindt, C.
\newblock Lee-Yang zeros and large-deviation statistics of a molecular zipper.
\newblock {\em Phys. Rev. E} {\bf 2018}, {\em 97},~012115.
\newblock {\url{https://doi.org/10.1103/physreve.97.012115}}.

\bibitem[Deger and Flindt(2020)]{Deger2020}
Deger, A.; Flindt, C.
\newblock Lee-{Y}ang theory of the {C}urie-{W}eiss model and its rare
  fluctuations.
\newblock {\em Phys. Rev. Research} {\bf 2020}, {\em 2},~033009.
\newblock {\url{https://doi.org/10.1103/physrevresearch.2.033009}}.

\bibitem[Deger et~al.(2020)Deger, Brange, and Flindt]{Deger2020a}
Deger, A.; Brange, F.; Flindt, C.
\newblock Lee-{Y}ang theory, high cumulants, and large-deviation statistics of
  the magnetization in the {I}sing model.
\newblock {\em Phys. Rev. B} {\bf 2020}, {\em 102},~174418.
\newblock {\url{https://doi.org/10.1103/physrevb.102.174418}}.

\bibitem[Janke and Kappler(1997)]{Janke1997}
Janke, W.; Kappler, S.
\newblock Monte-Carlo study of pure-phase cumulants of 2D $q$-state Potts
  models.
\newblock {\em J. Physique I} {\bf 1997}, {\em 7},~663--674.
\newblock {\url{https://doi.org/10.1051/jp1:1997183}}.

\bibitem[Gessert et~al.(2023)Gessert, Janke, and Weigel]{Gessert2023}
Gessert, D.; Janke, W.; Weigel, M.
\newblock Resampling schemes in population annealing -- numerical and
  theoretical results.
\newblock {\em Phys. Rev. E} {\bf 2023}, {\em 108},~065309.
\newblock {\url{https://doi.org/10.1103/PhysRevE.108.065309}}.

\bibitem[Weigel et~al.(2021)Weigel, Barash, Shchur, and Janke]{Weigel2021}
Weigel, M.; Barash, L.; Shchur, L.; Janke, W.
\newblock Understanding population annealing {M}onte {C}arlo simulations.
\newblock {\em Phys. Rev. E} {\bf 2021}, {\em 103},~053301.
\newblock {\url{https://doi.org/10.1103/physreve.103.053301}}.

\bibitem[Barash et~al.(2017)Barash, Weigel, Borovsk{\'{y}}, Janke, and
  Shchur]{Barash2017}
Barash, L.Y.; Weigel, M.; Borovsk{\'{y}}, M.; Janke, W.; Shchur, L.N.
\newblock {GPU} accelerated population annealing algorithm.
\newblock {\em Comput. Phys. Commun.} {\bf 2017}, {\em 220},~341--350.
\newblock {\url{https://doi.org/10.1016/j.cpc.2017.06.020}}.

\bibitem[Gessert et~al.()Gessert, Janke, and Weigel]{Gessert2024}
Gessert, D.; Janke, W.; Weigel, M.
\newblock (in preparation).

\bibitem[Levenberg(1944)]{Levenberg1944}
Levenberg, K.
\newblock A method for the solution of certain non-linear problems in least
  squares.
\newblock {\em Q. Appl. Math.} {\bf 1944}, {\em 2},~164--168.
\newblock {\url{https://doi.org/10.1090/qam/10666}}.

\bibitem[Marquardt(1963)]{Marquardt1963}
Marquardt, D.W.
\newblock An algorithm for least-squares estimation of nonlinear parameters.
\newblock {\em Journal of the Society for Industrial and Applied Mathematics}
  {\bf 1963}, {\em 11},~431--441.
\newblock {\url{https://doi.org/10.1137/0111030}}.

\bibitem[Ferrenberg and Landau(1991)]{ferrenberg:91a}
Ferrenberg, A.M.; Landau, D.P.
\newblock Critical behavior of the three-dimensional Ising model: A
  high-resolution Monte Carlo study.
\newblock {\em Phys. Rev. B} {\bf 1991}, {\em 44},~5081--5091.
\newblock {\url{https://doi.org/10.1103/PhysRevB.44.5081}}.

\bibitem[Press et~al.(2007)Press, Teukolsky, Vetterling, and
  Flannery]{Press2007}
Press, W.H.; Teukolsky, S.A.; Vetterling, W.T.; Flannery, B.P.
\newblock {\em Numerical Recipes}; Cambridge University Press: New York,  2007.

\bibitem[Efron(1982)]{efron1982}
Efron, B.
\newblock {\em The Jackknife, the Bootstrap, and other Resampling Plans};
  Society for Industrial and Applied Mathematics: Philadelphia, PA,  1982.
\newblock {\url{https://doi.org/10.1137/1.9781611970319}}.

\bibitem[Janke(2012)]{Janke2012}
Janke, W., Monte Carlo Simulations in Statistical Physics -- From Basic
  Principles to Advanced Applications.
\newblock In {\em Order, Disorder and Criticality}; edited by Holovatch, Y.;
  World Scientific: Singapore,  2012; pp. 93--166.
\newblock {\url{https://doi.org/10.1142/9789814417891_0003}}.

\bibitem[Matveev and Shrock(1996)]{Matveev1996}
Matveev, V.; Shrock, R.
\newblock Complex-temperature singularities in the $d=2$ {I}sing model:
  triangular and honeycomb lattices.
\newblock {\em J. Phys. A} {\bf 1996}, {\em 29},~803--823.
\newblock {\url{https://doi.org/10.1088/0305-4470/29/4/009}}.

\bibitem[Kim et~al.(2008)Kim, Hwang, and Kim]{Kim2008b}
Kim, S.Y.; Hwang, C.O.; Kim, J.M.
\newblock Partition function zeros of the antiferromagnetic Ising model on
  triangular lattice in the complex temperature plane for nonzero magnetic
  field.
\newblock {\em Nucl. Phys. B} {\bf 2008}, {\em 805},~441--450.
\newblock {\url{https://doi.org/10.1016/j.nuclphysb.2008.06.018}}.

\end{thebibliography}
\end{document}


\hypersetup{
  breaklinks=true,   
  colorlinks=true,   
  pdfusetitle=true,  
}

\tableofcontents

\section{Video description}
The supplementary video file shows the Fisher zeros of the $L = 4$ system for $\J2$ continuously varying from $0$ to $-1$ as determined by numerically solving for the roots of the polynomial, analogous to Figure~1 of the main text. Note that our Mathematica script fails to find zeros very close to the real axis. Hence, throughout the animation some zeros appear and disappear. This includes the leading zero which disappears for $\J2$ close to $-1/4$. By numerically evaluating Equation~(5a) of the main text, we confirmed that the zero does not actually disappear, but rather approaches the origin as $\J2$ goes to $-1/4$. 
\section{All FSS fitting results}
In this section we provide tables with the details of all performed FSS fits for varying fit intervals from $L_{\min}$ to $L_{\max}$. In all tables $\chi_\nu^2$ denotes the $\chi^2$  value per degree of freedom computed from the error-weighted fits.
\subsection{Temperature exponent $y_t$}%
\subsubsection{From the imaginary part of the Fisher zeros}
\begin{table}[H]
    \caption{Fitting parameters of FSS fits using the ansatz $\Im(\beta_0(L)) = a L^{-y_t}$ for $\J2=-0.1$ and different fitting ranges.\label{tab:fss_im_-0.1}}
    	\begin{adjustwidth}{-\extralength}{0cm}
	\begin{tabularx}{\fulllength}{llllll}
\toprule
$\mathbf{L_{\max}}$ & 24 & 32 & 48 & 64 & 88 \\
$\mathbf{L_{\min}}$ &  &  &  &  &  \\
\midrule
8 & \makecell[l]{$a = 1.249(17)$\\$y_t = 0.9668(61)$\\$\chi_\nu^2 = 0.26$} & \makecell[l]{$a = 1.249(14)$\\$y_t = 0.9667(48)$\\$\chi_\nu^2 = 0.13$} & \makecell[l]{$a = 1.285(10)$\\$y_t = 0.9796(31)$\\$\chi_\nu^2 = 4.25$} & \makecell[l]{$a = 1.2908(99)$\\$y_t = 0.9817(29)$\\$\chi_\nu^2 = 3.97$} & \makecell[l]{$a = 1.2942(98)$\\$y_t = 0.9829(29)$\\$\chi_\nu^2 = 4.21$} \\
16 & -- & \makecell[l]{$a = 1.269(60)$\\$y_t = 0.972(15)$\\$\chi_\nu^2 = 0.13$} & \makecell[l]{$a = 1.380(37)$\\$y_t = 1.0001(79)$\\$\chi_\nu^2 = 2.34$} & \makecell[l]{$a = 1.391(33)$\\$y_t = 1.0026(70)$\\$\chi_\nu^2 = 1.72$} & \makecell[l]{$a = 1.402(33)$\\$y_t = 1.0052(68)$\\$\chi_\nu^2 = 2.00$} \\
24 & -- & -- & \makecell[l]{$a = 1.467(71)$\\$y_t = 1.017(13)$\\$\chi_\nu^2 = 2.35$} & \makecell[l]{$a = 1.472(61)$\\$y_t = 1.018(11)$\\$\chi_\nu^2 = 1.18$} & \makecell[l]{$a = 1.495(60)$\\$y_t = 1.022(11)$\\$\chi_\nu^2 = 1.38$} \\
32 & -- & -- & -- & \makecell[l]{$a = 1.62(13)$\\$y_t = 1.042(21)$\\$\chi_\nu^2 = 0.36$} & \makecell[l]{$a = 1.66(12)$\\$y_t = 1.049(19)$\\$\chi_\nu^2 = 0.57$} \\
48 & -- & -- & -- & -- & \makecell[l]{$a = 1.65(25)$\\$y_t = 1.048(38)$\\$\chi_\nu^2 = 1.14$} \\
\bottomrule
\end{tabularx}

	\end{adjustwidth}
\end{table}
\begin{table}[H]
    \caption{As Table~\ref{tab:fss_im_-0.1} but for $\J2=-0.2$.\label{tab:fss_im_-0.2}}
    	\begin{adjustwidth}{-\extralength}{0cm}
	\begin{tabularx}{\fulllength}{llllll}
\toprule
$\mathbf{L_{\max}}$ & 24 & 32 & 48 & 64 & 88 \\
$\mathbf{L_{\min}}$ &  &  &  &  &  \\
\midrule
8 & \makecell[l]{$a = 2.856(25)$\\$y_t = 0.9215(35)$\\$\chi_\nu^2 = 1.72$} & \makecell[l]{$a = 2.888(23)$\\$y_t = 0.9263(31)$\\$\chi_\nu^2 = 5.25$} & \makecell[l]{$a = 2.960(20)$\\$y_t = 0.9369(26)$\\$\chi_\nu^2 = 15.93$} & \makecell[l]{$a = 2.982(18)$\\$y_t = 0.9400(23)$\\$\chi_\nu^2 = 13.61$} & \makecell[l]{$a = 3.021(17)$\\$y_t = 0.9453(21)$\\$\chi_\nu^2 = 17.73$} \\
16 & -- & \makecell[l]{$a = 3.097(77)$\\$y_t = 0.9500(86)$\\$\chi_\nu^2 = 1.58$} & \makecell[l]{$a = 3.244(52)$\\$y_t = 0.9663(54)$\\$\chi_\nu^2 = 3.78$} & \makecell[l]{$a = 3.221(42)$\\$y_t = 0.9639(42)$\\$\chi_\nu^2 = 2.71$} & \makecell[l]{$a = 3.281(37)$\\$y_t = 0.9702(37)$\\$\chi_\nu^2 = 4.17$} \\
24 & -- & -- & \makecell[l]{$a = 3.59(14)$\\$y_t = 0.994(12)$\\$\chi_\nu^2 = 0.13$} & \makecell[l]{$a = 3.40(10)$\\$y_t = 0.9780(83)$\\$\chi_\nu^2 = 2.13$} & \makecell[l]{$a = 3.502(86)$\\$y_t = 0.9872(68)$\\$\chi_\nu^2 = 2.58$} \\
32 & -- & -- & -- & \makecell[l]{$a = 3.26(19)$\\$y_t = 0.968(15)$\\$\chi_\nu^2 = 3.58$} & \makecell[l]{$a = 3.52(15)$\\$y_t = 0.988(11)$\\$\chi_\nu^2 = 3.87$} \\
48 & -- & -- & -- & -- & \makecell[l]{$a = 3.54(29)$\\$y_t = 0.990(20)$\\$\chi_\nu^2 = 7.72$} \\
\bottomrule
\end{tabularx}

	\end{adjustwidth}
\end{table}
\begin{table}[H]
    \caption{As Table~\ref{tab:fss_im_-0.1} but for $\J2=-0.21$.\label{tab:fss_im_-0.21}}
    	\begin{adjustwidth}{-\extralength}{0cm}
	\begin{tabularx}{\fulllength}{llllll}
\toprule
$\mathbf{L_{\max}}$ & 24 & 32 & 48 & 64 & 88 \\
$\mathbf{L_{\min}}$ &  &  &  &  &  \\
\midrule
8 & \makecell[l]{$a = 3.609(24)$\\$y_t = 0.9238(28)$\\$\chi_\nu^2 = 5.52$} & \makecell[l]{$a = 3.628(21)$\\$y_t = 0.9262(24)$\\$\chi_\nu^2 = 4.11$} & \makecell[l]{$a = 3.663(18)$\\$y_t = 0.9304(19)$\\$\chi_\nu^2 = 5.58$} & \makecell[l]{$a = 3.693(16)$\\$y_t = 0.9340(17)$\\$\chi_\nu^2 = 9.27$} & \makecell[l]{$a = 3.708(16)$\\$y_t = 0.9357(16)$\\$\chi_\nu^2 = 9.75$} \\
16 & -- & \makecell[l]{$a = 3.859(86)$\\$y_t = 0.9461(73)$\\$\chi_\nu^2 = 0.03$} & \makecell[l]{$a = 3.870(55)$\\$y_t = 0.9470(45)$\\$\chi_\nu^2 = 0.03$} & \makecell[l]{$a = 3.937(48)$\\$y_t = 0.9527(38)$\\$\chi_\nu^2 = 1.92$} & \makecell[l]{$a = 3.965(45)$\\$y_t = 0.9550(35)$\\$\chi_\nu^2 = 2.05$} \\
24 & -- & -- & \makecell[l]{$a = 3.87(10)$\\$y_t = 0.9471(77)$\\$\chi_\nu^2 = 0.05$} & \makecell[l]{$a = 4.008(85)$\\$y_t = 0.9575(61)$\\$\chi_\nu^2 = 2.36$} & \makecell[l]{$a = 4.053(77)$\\$y_t = 0.9609(53)$\\$\chi_\nu^2 = 2.03$} \\
32 & -- & -- & -- & \makecell[l]{$a = 4.21(18)$\\$y_t = 0.970(11)$\\$\chi_\nu^2 = 3.00$} & \makecell[l]{$a = 4.26(15)$\\$y_t = 0.9737(93)$\\$\chi_\nu^2 = 1.64$} \\
48 & -- & -- & -- & -- & \makecell[l]{$a = 4.70(33)$\\$y_t = 0.998(17)$\\$\chi_\nu^2 = 0.59$} \\
\bottomrule
\end{tabularx}

	\end{adjustwidth}
\end{table}
\begin{table}[H]
    \caption{As Table~\ref{tab:fss_im_-0.1} but for $\J2=-0.22$.\label{tab:fss_im_-0.22}}
    	\begin{adjustwidth}{-\extralength}{0cm}
	\begin{tabularx}{\fulllength}{llllll}
\toprule
$\mathbf{L_{\max}}$ & 24 & 32 & 48 & 64 & 88 \\
$\mathbf{L_{\min}}$ &  &  &  &  &  \\
\midrule
8 & \makecell[l]{$a = 5.088(34)$\\$y_t = 0.9328(30)$\\$\chi_\nu^2 = 3.39$} & \makecell[l]{$a = 5.118(28)$\\$y_t = 0.9356(23)$\\$\chi_\nu^2 = 2.77$} & \makecell[l]{$a = 5.128(26)$\\$y_t = 0.9364(21)$\\$\chi_\nu^2 = 2.15$} & \makecell[l]{$a = 5.179(23)$\\$y_t = 0.9410(18)$\\$\chi_\nu^2 = 5.59$} & \makecell[l]{$a = 5.239(20)$\\$y_t = 0.9461(15)$\\$\chi_\nu^2 = 9.45$} \\
16 & -- & \makecell[l]{$a = 5.40(13)$\\$y_t = 0.9525(76)$\\$\chi_\nu^2 = 0.04$} & \makecell[l]{$a = 5.38(11)$\\$y_t = 0.9513(63)$\\$\chi_\nu^2 = 0.06$} & \makecell[l]{$a = 5.503(80)$\\$y_t = 0.9588(45)$\\$\chi_\nu^2 = 1.00$} & \makecell[l]{$a = 5.597(62)$\\$y_t = 0.9644(32)$\\$\chi_\nu^2 = 1.56$} \\
24 & -- & -- & \makecell[l]{$a = 5.31(22)$\\$y_t = 0.948(13)$\\$\chi_\nu^2 = 0.01$} & \makecell[l]{$a = 5.60(14)$\\$y_t = 0.9634(71)$\\$\chi_\nu^2 = 1.16$} & \makecell[l]{$a = 5.714(97)$\\$y_t = 0.9697(46)$\\$\chi_\nu^2 = 1.24$} \\
32 & -- & -- & -- & \makecell[l]{$a = 5.77(24)$\\$y_t = 0.971(11)$\\$\chi_\nu^2 = 1.37$} & \makecell[l]{$a = 5.88(16)$\\$y_t = 0.9766(68)$\\$\chi_\nu^2 = 0.87$} \\
48 & -- & -- & -- & -- & \makecell[l]{$a = 6.35(45)$\\$y_t = 0.995(16)$\\$\chi_\nu^2 = 0.32$} \\
\bottomrule
\end{tabularx}

	\end{adjustwidth}
\end{table}

\subsubsection{From ordinary FSS}
\begin{table}[H]
    \caption{Fitting parameters of FSS fits using the ansatz $\dlm_{\max}(L) = a L^{y_t}$ for $\J2=-0.1$ and different fitting ranges.\label{tab:fss_dlm_-0.1}}
	\begin{adjustwidth}{-\extralength}{0cm}
	\begin{tabularx}{\fulllength}{llllll}
\toprule
$\mathbf{L_{\max}}$ & 24 & 32 & 48 & 64 & 88 \\
$\mathbf{L_{\min}}$ &  &  &  &  &  \\
\midrule
8 & \makecell[l]{$a = 0.6599(42)$\\$y_t = 0.9957(24)$\\$\chi_\nu^2 = 0.99$} & \makecell[l]{$a = 0.6619(33)$\\$y_t = 0.9943(17)$\\$\chi_\nu^2 = 0.82$} & \makecell[l]{$a = 0.6623(30)$\\$y_t = 0.9941(15)$\\$\chi_\nu^2 = 0.57$} & \makecell[l]{$a = 0.6627(28)$\\$y_t = 0.9939(13)$\\$\chi_\nu^2 = 0.45$} & \makecell[l]{$a = 0.6624(24)$\\$y_t = 0.9941(11)$\\$\chi_\nu^2 = 0.37$} \\
16 & -- & \makecell[l]{$a = 0.679(22)$\\$y_t = 0.9867(96)$\\$\chi_\nu^2 = 0.98$} & \makecell[l]{$a = 0.670(12)$\\$y_t = 0.9908(51)$\\$\chi_\nu^2 = 0.62$} & \makecell[l]{$a = 0.6681(82)$\\$y_t = 0.9916(35)$\\$\chi_\nu^2 = 0.43$} & \makecell[l]{$a = 0.6644(57)$\\$y_t = 0.9932(23)$\\$\chi_\nu^2 = 0.42$} \\
24 & -- & -- & \makecell[l]{$a = 0.670(12)$\\$y_t = 0.9908(51)$\\$\chi_\nu^2 = 0.25$} & \makecell[l]{$a = 0.6681(82)$\\$y_t = 0.9916(35)$\\$\chi_\nu^2 = 0.15$} & \makecell[l]{$a = 0.6644(57)$\\$y_t = 0.9933(23)$\\$\chi_\nu^2 = 0.23$} \\
32 & -- & -- & -- & \makecell[l]{$a = 0.663(12)$\\$y_t = 0.9935(50)$\\$\chi_\nu^2 = 0.01$} & \makecell[l]{$a = 0.6602(78)$\\$y_t = 0.9948(31)$\\$\chi_\nu^2 = 0.05$} \\
48 & -- & -- & -- & -- & \makecell[l]{$a = 0.657(18)$\\$y_t = 0.9959(65)$\\$\chi_\nu^2 = 0.07$} \\
\bottomrule
\end{tabularx}

	\end{adjustwidth}
\end{table}
\begin{table}[H]
    \caption{As Table~\ref{tab:fss_dlm_-0.1} but for $\J2=-0.2$.\label{tab:fss_dlm_-0.2}}
	\begin{adjustwidth}{-\extralength}{0cm}
	\begin{tabularx}{\fulllength}{llllll}
\toprule
$\mathbf{L_{\max}}$ & 24 & 32 & 48 & 64 & 88 \\
$\mathbf{L_{\min}}$ &  &  &  &  &  \\
\midrule
8 & \makecell[l]{$a = 0.3174(21)$\\$y_t = 0.9211(27)$\\$\chi_\nu^2 = 9.65$} & \makecell[l]{$a = 0.3123(15)$\\$y_t = 0.9284(18)$\\$\chi_\nu^2 = 11.55$} & \makecell[l]{$a = 0.3078(13)$\\$y_t = 0.9344(15)$\\$\chi_\nu^2 = 21.90$} & \makecell[l]{$a = 0.3051(12)$\\$y_t = 0.9380(14)$\\$\chi_\nu^2 = 25.60$} & \makecell[l]{$a = 0.29727(97)$\\$y_t = 0.9484(10)$\\$\chi_\nu^2 = 47.13$} \\
16 & -- & \makecell[l]{$a = 0.2918(44)$\\$y_t = 0.9493(47)$\\$\chi_\nu^2 = 0.00$} & \makecell[l]{$a = 0.2841(32)$\\$y_t = 0.9581(34)$\\$\chi_\nu^2 = 3.77$} & \makecell[l]{$a = 0.2803(27)$\\$y_t = 0.9625(29)$\\$\chi_\nu^2 = 4.35$} & \makecell[l]{$a = 0.2730(18)$\\$y_t = 0.9707(19)$\\$\chi_\nu^2 = 6.60$} \\
24 & -- & -- & \makecell[l]{$a = 0.2729(61)$\\$y_t = 0.9694(64)$\\$\chi_\nu^2 = 3.15$} & \makecell[l]{$a = 0.2689(46)$\\$y_t = 0.9737(48)$\\$\chi_\nu^2 = 2.10$} & \makecell[l]{$a = 0.2643(26)$\\$y_t = 0.9788(26)$\\$\chi_\nu^2 = 1.92$} \\
32 & -- & -- & -- & \makecell[l]{$a = 0.2597(62)$\\$y_t = 0.9829(65)$\\$\chi_\nu^2 = 0.00$} & \makecell[l]{$a = 0.2593(33)$\\$y_t = 0.9833(32)$\\$\chi_\nu^2 = 0.00$} \\
48 & -- & -- & -- & -- & \makecell[l]{$a = 0.2588(73)$\\$y_t = 0.9838(67)$\\$\chi_\nu^2 = 0.00$} \\
\bottomrule
\end{tabularx}

	\end{adjustwidth}
\end{table}
\begin{table}[H]
    \caption{As Table~\ref{tab:fss_dlm_-0.1} but for $\J2=-0.21$.\label{tab:fss_dlm_-0.21}}
	\begin{adjustwidth}{-\extralength}{0cm}
    \begin{tabularx}{\fulllength}{llllll}
\toprule
$\mathbf{L_{\max}}$ & 24 & 32 & 48 & 64 & 88 \\
$\mathbf{L_{\min}}$ &  &  &  &  &  \\
\midrule
8 & \makecell[l]{$a = 0.2733(23)$\\$y_t = 0.8953(31)$\\$\chi_\nu^2 = 15.57$} & \makecell[l]{$a = 0.2590(15)$\\$y_t = 0.9170(19)$\\$\chi_\nu^2 = 46.27$} & \makecell[l]{$a = 0.2505(11)$\\$y_t = 0.9294(14)$\\$\chi_\nu^2 = 63.30$} & \makecell[l]{$a = 0.2484(11)$\\$y_t = 0.9323(13)$\\$\chi_\nu^2 = 55.68$} & \makecell[l]{$a = 0.24342(96)$\\$y_t = 0.9392(12)$\\$\chi_\nu^2 = 69.06$} \\
16 & -- & \makecell[l]{$a = 0.2401(24)$\\$y_t = 0.9405(31)$\\$\chi_\nu^2 = 4.61$} & \makecell[l]{$a = 0.2342(16)$\\$y_t = 0.9488(20)$\\$\chi_\nu^2 = 8.41$} & \makecell[l]{$a = 0.2328(15)$\\$y_t = 0.9508(19)$\\$\chi_\nu^2 = 7.25$} & \makecell[l]{$a = 0.2284(12)$\\$y_t = 0.9567(16)$\\$\chi_\nu^2 = 13.99$} \\
24 & -- & -- & \makecell[l]{$a = 0.2214(34)$\\$y_t = 0.9642(43)$\\$\chi_\nu^2 = 0.00$} & \makecell[l]{$a = 0.2209(28)$\\$y_t = 0.9648(35)$\\$\chi_\nu^2 = 0.03$} & \makecell[l]{$a = 0.2161(21)$\\$y_t = 0.9711(26)$\\$\chi_\nu^2 = 2.21$} \\
32 & -- & -- & -- & \makecell[l]{$a = 0.2206(36)$\\$y_t = 0.9652(44)$\\$\chi_\nu^2 = 0.05$} & \makecell[l]{$a = 0.2146(24)$\\$y_t = 0.9727(30)$\\$\chi_\nu^2 = 2.72$} \\
48 & -- & -- & -- & -- & \makecell[l]{$a = 0.2064(46)$\\$y_t = 0.9822(55)$\\$\chi_\nu^2 = 1.33$} \\
\bottomrule
\end{tabularx}

	\end{adjustwidth}
\end{table}
\begin{table}[H]
    \caption{As Table~\ref{tab:fss_dlm_-0.1} but for $\J2=-0.22$.\label{tab:fss_dlm_-0.22}}
	\begin{adjustwidth}{-\extralength}{0cm}
	\begin{tabularx}{\fulllength}{llllll}
\toprule
$\mathbf{L_{\max}}$ & 24 & 32 & 48 & 64 & 88 \\
$\mathbf{L_{\min}}$ &  &  &  &  &  \\
\midrule
8 & \makecell[l]{$a = 0.2122(14)$\\$y_t = 0.8781(25)$\\$\chi_\nu^2 = 50.78$} & \makecell[l]{$a = 0.2060(11)$\\$y_t = 0.8907(19)$\\$\chi_\nu^2 = 54.35$} & \makecell[l]{$a = 0.20043(87)$\\$y_t = 0.9019(15)$\\$\chi_\nu^2 = 70.52$} & \makecell[l]{$a = 0.19573(76)$\\$y_t = 0.9112(13)$\\$\chi_\nu^2 = 86.54$} & \makecell[l]{$a = 0.19291(71)$\\$y_t = 0.9167(12)$\\$\chi_\nu^2 = 95.88$} \\
16 & -- & \makecell[l]{$a = 0.1819(24)$\\$y_t = 0.9302(42)$\\$\chi_\nu^2 = 0.00$} & \makecell[l]{$a = 0.1785(16)$\\$y_t = 0.9365(28)$\\$\chi_\nu^2 = 2.01$} & \makecell[l]{$a = 0.1750(13)$\\$y_t = 0.9431(22)$\\$\chi_\nu^2 = 5.70$} & \makecell[l]{$a = 0.1721(11)$\\$y_t = 0.9484(19)$\\$\chi_\nu^2 = 11.45$} \\
24 & -- & -- & \makecell[l]{$a = 0.1740(33)$\\$y_t = 0.9437(54)$\\$\chi_\nu^2 = 1.53$} & \makecell[l]{$a = 0.1689(22)$\\$y_t = 0.9525(36)$\\$\chi_\nu^2 = 3.22$} & \makecell[l]{$a = 0.1645(18)$\\$y_t = 0.9600(30)$\\$\chi_\nu^2 = 7.05$} \\
32 & -- & -- & -- & \makecell[l]{$a = 0.1632(33)$\\$y_t = 0.9612(53)$\\$\chi_\nu^2 = 1.47$} & \makecell[l]{$a = 0.1580(26)$\\$y_t = 0.9702(42)$\\$\chi_\nu^2 = 4.60$} \\
48 & -- & -- & -- & -- & \makecell[l]{$a = 0.1452(50)$\\$y_t = 0.9905(84)$\\$\chi_\nu^2 = 1.47$} \\
\bottomrule
\end{tabularx}

	\end{adjustwidth}
\end{table}

\subsection{Determining the inverse critical temperature $\beta_c$ from the real part of the Fisher zeros}%
With the limited number of data points, a three-parameter fit for the real part of the Fisher zero is rather unstable. 
Therefore, we fix $y_t=1$ in the fit ansatz $\Re(\beta_0(L)) = \beta_c - a L^{-y_t} - b L^{-2y_t}$. Results are shown in Tables~\ref{tab:fss_re_-0.1} to~\ref{tab:fss_re_-0.22}.
\begin{table}[H]
    \caption{Fitting parameters of FSS fits using the ansatz $\Re(\beta_0(L)) = \beta_c - a L^{-y_t} - b L^{-2y_t}$ ($y_t=1$ fixed) for $\J2=-0.1$ and different fitting ranges.\label{tab:fss_re_-0.1}}
    	\begin{adjustwidth}{-\extralength}{0cm}
	\begin{tabularx}{\fulllength}{lllll}
\toprule
$\mathbf{L_{\max}}$ & 32 & 48 & 64 & 88 \\
$\mathbf{L_{\min}}$ &  &  &  &  \\
\midrule
8 & \makecell[l]{$a = 0.279(41)$\\$b = 1.13(25)$\\$\beta_c = 1.0226(13)/ J_1$\\$\chi_\nu^2 = 0.29$} & \makecell[l]{$a = 0.310(21)$\\$b = 0.95(15)$\\$\beta_c = 1.02371(52)/ J_1$\\$\chi_\nu^2 = 0.55$} & \makecell[l]{$a = 0.301(19)$\\$b = 1.00(14)$\\$\beta_c = 1.02344(43)/ J_1$\\$\chi_\nu^2 = 0.65$} & \makecell[l]{$a = 0.304(16)$\\$b = 0.99(12)$\\$\beta_c = 1.02350(32)/ J_1$\\$\chi_\nu^2 = 0.50$} \\
16 & -- & \makecell[l]{$a = 0.351(75)$\\$b = 0.44(92)$\\$\beta_c = 1.0244(13)/ J_1$\\$\chi_\nu^2 = 0.79$} & \makecell[l]{$a = 0.300(59)$\\$b = 1.02(75)$\\$\beta_c = 1.02342(94)/ J_1$\\$\chi_\nu^2 = 0.98$} & \makecell[l]{$a = 0.309(41)$\\$b = 0.92(55)$\\$\beta_c = 1.02357(58)/ J_1$\\$\chi_\nu^2 = 0.66$} \\
24 & -- & -- & \makecell[l]{$a = 0.27(15)$\\$b = 1.6(2.6)$\\$\beta_c = 1.0230(20)/ J_1$\\$\chi_\nu^2 = 1.89$} & \makecell[l]{$a = 0.307(79)$\\$b = 1.0(1.5)$\\$\beta_c = 1.02355(91)/ J_1$\\$\chi_\nu^2 = 1.00$} \\
32 & -- & -- & -- & \makecell[l]{$a = 0.22(14)$\\$b = 3.4(3.3)$\\$\beta_c = 1.0228(13)/ J_1$\\$\chi_\nu^2 = 1.29$} \\
\bottomrule
\end{tabularx}

	\end{adjustwidth}
\end{table}
\begin{table}[H]
    \caption{As Table~\ref{tab:fss_re_-0.1} but for $\J2=-0.2$.\label{tab:fss_re_-0.2}}
    	\begin{adjustwidth}{-\extralength}{0cm}
	\begin{tabularx}{\fulllength}{lllll}
\toprule
$\mathbf{L_{\max}}$ & 32 & 48 & 64 & 88 \\
$\mathbf{L_{\min}}$ &  &  &  &  \\
\midrule
8 & \makecell[l]{$a = 0.813(62)$\\$b = 6.64(38)$\\$\beta_c = 2.6563(20)/ J_1$\\$\chi_\nu^2 = 0.01$} & \makecell[l]{$a = 0.803(34)$\\$b = 6.70(23)$\\$\beta_c = 2.65588(85)/ J_1$\\$\chi_\nu^2 = 0.02$} & \makecell[l]{$a = 0.841(30)$\\$b = 6.47(21)$\\$\beta_c = 2.65706(72)/ J_1$\\$\chi_\nu^2 = 2.22$} & \makecell[l]{$a = 0.853(25)$\\$b = 6.40(18)$\\$\beta_c = 2.65740(54)/ J_1$\\$\chi_\nu^2 = 1.79$} \\
16 & -- & \makecell[l]{$a = 0.79(11)$\\$b = 6.9(1.4)$\\$\beta_c = 2.6556(20)/ J_1$\\$\chi_\nu^2 = 0.03$} & \makecell[l]{$a = 0.940(92)$\\$b = 5.2(1.2)$\\$\beta_c = 2.6586(15)/ J_1$\\$\chi_\nu^2 = 2.68$} & \makecell[l]{$a = 0.932(64)$\\$b = 5.26(87)$\\$\beta_c = 2.65843(94)/ J_1$\\$\chi_\nu^2 = 1.79$} \\
24 & -- & -- & \makecell[l]{$a = 1.29(25)$\\$b = -1.0(4.2)$\\$\beta_c = 2.6630(33)/ J_1$\\$\chi_\nu^2 = 3.02$} & \makecell[l]{$a = 1.04(13)$\\$b = 3.1(2.4)$\\$\beta_c = 2.6596(15)/ J_1$\\$\chi_\nu^2 = 2.21$} \\
32 & -- & -- & -- & \makecell[l]{$a = 1.12(22)$\\$b = 1.1(5.3)$\\$\beta_c = 2.6603(22)/ J_1$\\$\chi_\nu^2 = 4.23$} \\
\bottomrule
\end{tabularx}

	\end{adjustwidth}
\end{table}
\begin{table}[H]
    \caption{As Table~\ref{tab:fss_re_-0.1} but for $\J2=-0.21$.\label{tab:fss_re_-0.21}}
	\begin{adjustwidth}{-\extralength}{0cm}
	\begin{tabularx}{\fulllength}{lllll}
\toprule
$\mathbf{L_{\max}}$ & 32 & 48 & 64 & 88 \\
$\mathbf{L_{\min}}$ &  &  &  &  \\
\midrule
8 & \makecell[l]{$a = 1.010(65)$\\$b = 9.91(39)$\\$\beta_c = 3.2539(23)/ J_1$\\$\chi_\nu^2 = 4.80$} & \makecell[l]{$a = 1.040(31)$\\$b = 9.74(22)$\\$\beta_c = 3.25504(91)/ J_1$\\$\chi_\nu^2 = 2.54$} & \makecell[l]{$a = 1.056(25)$\\$b = 9.65(19)$\\$\beta_c = 3.25560(64)/ J_1$\\$\chi_\nu^2 = 1.94$} & \makecell[l]{$a = 1.106(20)$\\$b = 9.34(17)$\\$\beta_c = 3.25721(42)/ J_1$\\$\chi_\nu^2 = 4.22$} \\
16 & -- & \makecell[l]{$a = 1.22(13)$\\$b = 7.6(1.6)$\\$\beta_c = 3.2581(24)/ J_1$\\$\chi_\nu^2 = 3.13$} & \makecell[l]{$a = 1.208(97)$\\$b = 7.7(1.2)$\\$\beta_c = 3.2578(15)/ J_1$\\$\chi_\nu^2 = 1.58$} & \makecell[l]{$a = 1.314(64)$\\$b = 6.48(86)$\\$\beta_c = 3.25965(83)/ J_1$\\$\chi_\nu^2 = 1.76$} \\
24 & -- & -- & \makecell[l]{$a = 0.94(27)$\\$b = 12.6(4.7)$\\$\beta_c = 3.2545(34)/ J_1$\\$\chi_\nu^2 = 2.00$} & \makecell[l]{$a = 1.36(13)$\\$b = 5.6(2.6)$\\$\beta_c = 3.2601(14)/ J_1$\\$\chi_\nu^2 = 2.57$} \\
32 & -- & -- & -- & \makecell[l]{$a = 1.81(24)$\\$b = -7.3(6.3)$\\$\beta_c = 3.2637(22)/ J_1$\\$\chi_\nu^2 = 0.18$} \\
\bottomrule
\end{tabularx}

	\end{adjustwidth}
\end{table}
\begin{table}[H]
    \caption{As Table~\ref{tab:fss_re_-0.1} but for $\J2=-0.22$.\label{tab:fss_re_-0.22}}
	\begin{adjustwidth}{-\extralength}{0cm}
	\begin{tabularx}{\fulllength}{lllll}
\toprule
$\mathbf{L_{\max}}$ & 32 & 48 & 64 & 88 \\
$\mathbf{L_{\min}}$ &  &  &  &  \\
\midrule
8 & \makecell[l]{$a = 1.565(76)$\\$b = 13.85(50)$\\$\beta_c = 4.2722(22)/ J_1$\\$\chi_\nu^2 = 1.14$} & \makecell[l]{$a = 1.481(44)$\\$b = 14.36(33)$\\$\beta_c = 4.26966(98)/ J_1$\\$\chi_\nu^2 = 1.47$} & \makecell[l]{$a = 1.511(37)$\\$b = 14.17(29)$\\$\beta_c = 4.27045(77)/ J_1$\\$\chi_\nu^2 = 1.55$} & \makecell[l]{$a = 1.500(31)$\\$b = 14.25(25)$\\$\beta_c = 4.27017(59)/ J_1$\\$\chi_\nu^2 = 1.24$} \\
16 & -- & \makecell[l]{$a = 1.40(12)$\\$b = 15.5(1.5)$\\$\beta_c = 4.2683(20)/ J_1$\\$\chi_\nu^2 = 2.39$} & \makecell[l]{$a = 1.518(90)$\\$b = 14.1(1.2)$\\$\beta_c = 4.2705(14)/ J_1$\\$\chi_\nu^2 = 2.32$} & \makecell[l]{$a = 1.486(65)$\\$b = 14.46(93)$\\$\beta_c = 4.27000(94)/ J_1$\\$\chi_\nu^2 = 1.63$} \\
24 & -- & -- & \makecell[l]{$a = 1.49(21)$\\$b = 14.6(3.8)$\\$\beta_c = 4.2702(27)/ J_1$\\$\chi_\nu^2 = 4.62$} & \makecell[l]{$a = 1.44(13)$\\$b = 15.5(2.5)$\\$\beta_c = 4.2695(15)/ J_1$\\$\chi_\nu^2 = 2.36$} \\
32 & -- & -- & -- & \makecell[l]{$a = 1.65(20)$\\$b = 10.2(4.6)$\\$\beta_c = 4.2715(21)/ J_1$\\$\chi_\nu^2 = 2.84$} \\
\bottomrule
\end{tabularx}

	\end{adjustwidth}
\end{table}

\subsection{Field exponent $y_h$}%
We use either the imaginary part of the Lee-Yang zeros or the magnetic susceptibility to obtain $y_h$, in both cases considering the system directly at the infinite-volume inverse critical temperature $\beta_c$. As described in the main text, the error bars for $y_h$ obtained after fixing $\beta_c$ at the values of Tables~\ref{tab:fss_re_-0.1}-\ref{tab:fss_re_-0.22} using FSS fits do not account for the uncertainty in $\beta_c$. Therefore, we use jackknifing over the whole process to correctly include both the statistical error in $\beta_c$ as well as that in the Lee-Yang zeros and the susceptibility, respectively.
\subsubsection{From the imaginary part of the Lee-Yang zeros}
\begin{table}[H]
    \caption{Fitting parameters of FSS fits at $\beta_c$ using the ansatz $\Im(h_0(L)) = a L^{-y_h}$ for $\J2=-0.1$ and different fitting ranges, using the jackknife procedure described in the main text.\label{tab:fss_imH_-0.1}}
	\begin{adjustwidth}{-\extralength}{0cm}
	\begin{tabularx}{\fulllength}{llllll}
\toprule
$\mathbf{L_{\max}}$ & 24 & 32 & 48 & 64 & 88 \\
$\mathbf{L_{\min}}$ &  &  &  &  &  \\
\midrule
8 & \makecell[l]{$a = 0.7329(11)$\\$y_h = 1.87629(97)$\\$\chi_\nu^2 = 0.22(56)$} & \makecell[l]{$a = 0.73178(82)$\\$y_h = 1.87564(69)$\\$\chi_\nu^2 = 1.6(1.7)$} & \makecell[l]{$a = 0.7317(17)$\\$y_h = 1.8756(13)$\\$\chi_\nu^2 = 1.4(1.9)$} & \makecell[l]{$a = 0.7313(28)$\\$y_h = 1.8754(20)$\\$\chi_\nu^2 = 1.5(2.6)$} & \makecell[l]{$a = 0.7313(38)$\\$y_h = 1.8754(25)$\\$\chi_\nu^2 = 1.5(3.1)$} \\
16 & -- & \makecell[l]{$a = 0.7297(26)$\\$y_h = 1.8747(13)$\\$\chi_\nu^2 = 2.3(1.8)$} & \makecell[l]{$a = 0.7305(41)$\\$y_h = 1.8751(23)$\\$\chi_\nu^2 = 1.6(3.0)$} & \makecell[l]{$a = 0.7302(55)$\\$y_h = 1.8750(30)$\\$\chi_\nu^2 = 1.2(3.0)$} & \makecell[l]{$a = 0.7306(67)$\\$y_h = 1.8751(35)$\\$\chi_\nu^2 = 1.1(3.4)$} \\
24 & -- & -- & \makecell[l]{$a = 0.7290(80)$\\$y_h = 1.8745(38)$\\$\chi_\nu^2 = 2.6(6.3)$} & \makecell[l]{$a = 0.7294(91)$\\$y_h = 1.8747(42)$\\$\chi_\nu^2 = 1.4(3.5)$} & \makecell[l]{$a = 0.730(10)$\\$y_h = 1.8750(46)$\\$\chi_\nu^2 = 1.1(2.6)$} \\
32 & -- & -- & -- & \makecell[l]{$a = 0.733(11)$\\$y_h = 1.8758(49)$\\$\chi_\nu^2 = 0.7(1.6)$} & \makecell[l]{$a = 0.733(11)$\\$y_h = 1.8757(50)$\\$\chi_\nu^2 = 0.42(88)$} \\
48 & -- & -- & -- & -- & \makecell[l]{$a = 0.731(11)$\\$y_h = 1.8752(49)$\\$\chi_\nu^2 = 0.4(1.3)$} \\
\bottomrule
\end{tabularx}

	\end{adjustwidth}
\end{table}
\begin{table}[H]
    \caption{As Table~\ref{tab:fss_imH_-0.1} but for $\J2=-0.2$.\label{tab:fss_imH_-0.2}}
	\begin{adjustwidth}{-\extralength}{0cm}
	\begin{tabularx}{\fulllength}{llllll}
\toprule
$\mathbf{L_{\max}}$ & 24 & 32 & 48 & 64 & 88 \\
$\mathbf{L_{\min}}$ &  &  &  &  &  \\
\midrule
8 & \makecell[l]{$a = 0.26075(58)$\\$y_h = 1.8780(13)$\\$\chi_\nu^2 = 0.8(4.3)$} & \makecell[l]{$a = 0.26028(72)$\\$y_h = 1.8772(16)$\\$\chi_\nu^2 = 2.1(4.1)$} & \makecell[l]{$a = 0.26021(83)$\\$y_h = 1.8771(18)$\\$\chi_\nu^2 = 1.5(3.1)$} & \makecell[l]{$a = 0.26013(95)$\\$y_h = 1.8769(20)$\\$\chi_\nu^2 = 1.5(3.6)$} & \makecell[l]{$a = 0.2600(15)$\\$y_h = 1.8767(30)$\\$\chi_\nu^2 = 1.9(4.7)$} \\
16 & -- & \makecell[l]{$a = 0.2584(16)$\\$y_h = 1.8750(26)$\\$\chi_\nu^2 = 0.6(2.0)$} & \makecell[l]{$a = 0.2590(17)$\\$y_h = 1.8757(28)$\\$\chi_\nu^2 = 0.62(87)$} & \makecell[l]{$a = 0.2590(21)$\\$y_h = 1.8757(32)$\\$\chi_\nu^2 = 0.47(65)$} & \makecell[l]{$a = 0.2592(30)$\\$y_h = 1.8759(44)$\\$\chi_\nu^2 = 0.51(73)$} \\
24 & -- & -- & \makecell[l]{$a = 0.2592(24)$\\$y_h = 1.8759(35)$\\$\chi_\nu^2 = 1.0(1.8)$} & \makecell[l]{$a = 0.2590(26)$\\$y_h = 1.8757(38)$\\$\chi_\nu^2 = 0.64(88)$} & \makecell[l]{$a = 0.2593(36)$\\$y_h = 1.8760(48)$\\$\chi_\nu^2 = 0.54(77)$} \\
32 & -- & -- & -- & \makecell[l]{$a = 0.2598(30)$\\$y_h = 1.8765(41)$\\$\chi_\nu^2 = 0.5(1.4)$} & \makecell[l]{$a = 0.2596(40)$\\$y_h = 1.8763(52)$\\$\chi_\nu^2 = 0.35(88)$} \\
48 & -- & -- & -- & -- & \makecell[l]{$a = 0.2591(55)$\\$y_h = 1.8758(66)$\\$\chi_\nu^2 = 0.3(1.4)$} \\
\bottomrule
\end{tabularx}

	\end{adjustwidth}
\end{table}
\begin{table}[H]
    \caption{As Table~\ref{tab:fss_imH_-0.1} but for $\J2=-0.21$.\label{tab:fss_imH_-0.21}}
	\begin{adjustwidth}{-\extralength}{0cm}
    \begin{tabularx}{\fulllength}{llllll}
\toprule
$\mathbf{L_{\max}}$ & 24 & 32 & 48 & 64 & 88 \\
$\mathbf{L_{\min}}$ &  &  &  &  &  \\
\midrule
8 & \makecell[l]{$a = 0.21298(68)$\\$y_h = 1.8826(16)$\\$\chi_\nu^2 = 0.17(52)$} & \makecell[l]{$a = 0.21264(67)$\\$y_h = 1.8819(17)$\\$\chi_\nu^2 = 1.8(2.1)$} & \makecell[l]{$a = 0.21235(79)$\\$y_h = 1.8813(20)$\\$\chi_\nu^2 = 1.7(2.0)$} & \makecell[l]{$a = 0.21240(90)$\\$y_h = 1.8814(22)$\\$\chi_\nu^2 = 1.4(1.4)$} & \makecell[l]{$a = 0.2125(14)$\\$y_h = 1.8817(30)$\\$\chi_\nu^2 = 1.5(2.0)$} \\
16 & -- & \makecell[l]{$a = 0.2114(14)$\\$y_h = 1.8799(27)$\\$\chi_\nu^2 = 1.6(2.2)$} & \makecell[l]{$a = 0.2116(13)$\\$y_h = 1.8803(26)$\\$\chi_\nu^2 = 0.9(1.2)$} & \makecell[l]{$a = 0.2119(14)$\\$y_h = 1.8807(29)$\\$\chi_\nu^2 = 1.0(1.4)$} & \makecell[l]{$a = 0.2124(20)$\\$y_h = 1.8814(37)$\\$\chi_\nu^2 = 1.5(2.3)$} \\
24 & -- & -- & \makecell[l]{$a = 0.2112(17)$\\$y_h = 1.8798(30)$\\$\chi_\nu^2 = 1.4(2.1)$} & \makecell[l]{$a = 0.2119(20)$\\$y_h = 1.8807(35)$\\$\chi_\nu^2 = 1.5(2.0)$} & \makecell[l]{$a = 0.2127(27)$\\$y_h = 1.8818(44)$\\$\chi_\nu^2 = 1.6(2.0)$} \\
32 & -- & -- & -- & \makecell[l]{$a = 0.2143(25)$\\$y_h = 1.8836(38)$\\$\chi_\nu^2 = 0.21(69)$} & \makecell[l]{$a = 0.2139(30)$\\$y_h = 1.8831(48)$\\$\chi_\nu^2 = 0.26(77)$} \\
48 & -- & -- & -- & -- & \makecell[l]{$a = 0.2139(39)$\\$y_h = 1.8832(57)$\\$\chi_\nu^2 = 0.2(1.1)$} \\
\bottomrule
\end{tabularx}

	\end{adjustwidth}
\end{table}
\begin{table}[H]
    \caption{As Table~\ref{tab:fss_imH_-0.1} but for $\J2=-0.22$.\label{tab:fss_imH_-0.22}}
	\begin{adjustwidth}{-\extralength}{0cm}
	\begin{tabularx}{\fulllength}{llllll}
\toprule
$\mathbf{L_{\max}}$ & 24 & 32 & 48 & 64 & 88 \\
$\mathbf{L_{\min}}$ &  &  &  &  &  \\
\midrule
8 & \makecell[l]{$a = 0.16181(28)$\\$y_h = 1.88294(89)$\\$\chi_\nu^2 = 1.1(1.8)$} & \makecell[l]{$a = 0.16140(38)$\\$y_h = 1.8818(11)$\\$\chi_\nu^2 = 3.7(2.9)$} & \makecell[l]{$a = 0.16102(42)$\\$y_h = 1.8808(12)$\\$\chi_\nu^2 = 6.2(3.5)$} & \makecell[l]{$a = 0.16083(36)$\\$y_h = 1.8804(10)$\\$\chi_\nu^2 = 5.3(2.5)$} & \makecell[l]{$a = 0.16081(36)$\\$y_h = 1.8803(10)$\\$\chi_\nu^2 = 4.3(1.9)$} \\
16 & -- & \makecell[l]{$a = 0.1592(11)$\\$y_h = 1.8777(22)$\\$\chi_\nu^2 = 1.1(2.2)$} & \makecell[l]{$a = 0.15869(85)$\\$y_h = 1.8766(16)$\\$\chi_\nu^2 = 1.0(1.4)$} & \makecell[l]{$a = 0.15944(62)$\\$y_h = 1.8781(14)$\\$\chi_\nu^2 = 1.6(1.7)$} & \makecell[l]{$a = 0.15948(58)$\\$y_h = 1.8781(13)$\\$\chi_\nu^2 = 1.2(1.3)$} \\
24 & -- & -- & \makecell[l]{$a = 0.15792(83)$\\$y_h = 1.8752(17)$\\$\chi_\nu^2 = 0.14(64)$} & \makecell[l]{$a = 0.15935(84)$\\$y_h = 1.8779(18)$\\$\chi_\nu^2 = 2.2(2.5)$} & \makecell[l]{$a = 0.15941(76)$\\$y_h = 1.8780(17)$\\$\chi_\nu^2 = 1.6(1.7)$} \\
32 & -- & -- & -- & \makecell[l]{$a = 0.1602(12)$\\$y_h = 1.8792(23)$\\$\chi_\nu^2 = 2.5(3.2)$} & \makecell[l]{$a = 0.1602(11)$\\$y_h = 1.8793(21)$\\$\chi_\nu^2 = 1.3(1.7)$} \\
48 & -- & -- & -- & -- & \makecell[l]{$a = 0.1625(22)$\\$y_h = 1.8827(36)$\\$\chi_\nu^2 = 0.7(1.9)$} \\
\bottomrule
\end{tabularx}

	\end{adjustwidth}
\end{table}
\subsubsection{From ordinary FSS}
\begin{table}[H]
    \caption{Fitting parameters of FSS fits at $\beta_c$ using the ansatz $\chi_L(\beta_c) = a L^{2 y_h -D}$ for $\J2=-0.1$ and different fitting ranges, using the jackknife procedure described in the main text. $D=2$ is the spatial dimension.\label{tab:fss_chi_-0.1}}
	\begin{adjustwidth}{-\extralength}{0cm}
	\begin{tabularx}{\fulllength}{llllll}
\toprule
$\mathbf{L_{\max}}$ & 24 & 32 & 48 & 64 & 88 \\
$\mathbf{L_{\min}}$ &  &  &  &  &  \\
\midrule
8 & \makecell[l]{$a = 2.3235(60)$\\$y_h = 1.87813(86)$\\$\chi_\nu^2 = 0.20(47)$} & \makecell[l]{$a = 2.3324(45)$\\$y_h = 1.87730(57)$\\$\chi_\nu^2 = 3.9(3.4)$} & \makecell[l]{$a = 2.3364(88)$\\$y_h = 1.8769(11)$\\$\chi_\nu^2 = 3.6(3.3)$} & \makecell[l]{$a = 2.342(16)$\\$y_h = 1.8764(17)$\\$\chi_\nu^2 = 4.5(5.5)$} & \makecell[l]{$a = 2.344(21)$\\$y_h = 1.8763(21)$\\$\chi_\nu^2 = 4.2(5.2)$} \\
16 & -- & \makecell[l]{$a = 2.359(14)$\\$y_h = 1.8755(11)$\\$\chi_\nu^2 = 2.7(2.0)$} & \makecell[l]{$a = 2.356(23)$\\$y_h = 1.8757(20)$\\$\chi_\nu^2 = 1.7(3.0)$} & \makecell[l]{$a = 2.360(31)$\\$y_h = 1.8754(26)$\\$\chi_\nu^2 = 1.4(2.5)$} & \makecell[l]{$a = 2.358(38)$\\$y_h = 1.8755(31)$\\$\chi_\nu^2 = 1.3(2.9)$} \\
24 & -- & -- & \makecell[l]{$a = 2.367(46)$\\$y_h = 1.8751(34)$\\$\chi_\nu^2 = 2.7(6.5)$} & \makecell[l]{$a = 2.367(51)$\\$y_h = 1.8750(37)$\\$\chi_\nu^2 = 1.4(3.4)$} & \makecell[l]{$a = 2.363(57)$\\$y_h = 1.8753(40)$\\$\chi_\nu^2 = 1.1(2.7)$} \\
32 & -- & -- & -- & \makecell[l]{$a = 2.351(62)$\\$y_h = 1.8759(42)$\\$\chi_\nu^2 = 1.0(1.9)$} & \makecell[l]{$a = 2.351(63)$\\$y_h = 1.8759(43)$\\$\chi_\nu^2 = 0.6(1.0)$} \\
48 & -- & -- & -- & -- & \makecell[l]{$a = 2.361(64)$\\$y_h = 1.8754(43)$\\$\chi_\nu^2 = 0.7(1.7)$} \\
\bottomrule
\end{tabularx}

	\end{adjustwidth}
\end{table}
\begin{table}[H]
    \caption{As Table~\ref{tab:fss_chi_-0.1} but for $\J2=-0.2$.\label{tab:fss_chi_-0.2}}
	\begin{adjustwidth}{-\extralength}{0cm}
	\begin{tabularx}{\fulllength}{llllll}
\toprule
$\mathbf{L_{\max}}$ & 24 & 32 & 48 & 64 & 88 \\
$\mathbf{L_{\min}}$ &  &  &  &  &  \\
\midrule
8 & \makecell[l]{$a = 6.944(31)$\\$y_h = 1.8828(13)$\\$\chi_\nu^2 = 5(13)$} & \makecell[l]{$a = 6.990(37)$\\$y_h = 1.8813(14)$\\$\chi_\nu^2 = 11(15)$} & \makecell[l]{$a = 7.013(43)$\\$y_h = 1.8805(16)$\\$\chi_\nu^2 = 11(14)$} & \makecell[l]{$a = 7.027(48)$\\$y_h = 1.8801(17)$\\$\chi_\nu^2 = 12(15)$} & \makecell[l]{$a = 7.067(85)$\\$y_h = 1.8789(28)$\\$\chi_\nu^2 = 18(23)$} \\
16 & -- & \makecell[l]{$a = 7.202(79)$\\$y_h = 1.8767(23)$\\$\chi_\nu^2 = 1.2(3.1)$} & \makecell[l]{$a = 7.200(86)$\\$y_h = 1.8767(24)$\\$\chi_\nu^2 = 0.7(1.6)$} & \makecell[l]{$a = 7.21(10)$\\$y_h = 1.8765(28)$\\$\chi_\nu^2 = 0.6(1.1)$} & \makecell[l]{$a = 7.22(15)$\\$y_h = 1.8763(38)$\\$\chi_\nu^2 = 0.6(1.1)$} \\
24 & -- & -- & \makecell[l]{$a = 7.23(12)$\\$y_h = 1.8761(30)$\\$\chi_\nu^2 = 0.8(1.5)$} & \makecell[l]{$a = 7.24(13)$\\$y_h = 1.8759(33)$\\$\chi_\nu^2 = 0.53(77)$} & \makecell[l]{$a = 7.23(18)$\\$y_h = 1.8761(42)$\\$\chi_\nu^2 = 0.46(67)$} \\
32 & -- & -- & -- & \makecell[l]{$a = 7.21(15)$\\$y_h = 1.8765(36)$\\$\chi_\nu^2 = 0.5(1.4)$} & \makecell[l]{$a = 7.22(20)$\\$y_h = 1.8763(46)$\\$\chi_\nu^2 = 0.36(88)$} \\
48 & -- & -- & -- & -- & \makecell[l]{$a = 7.24(27)$\\$y_h = 1.8759(57)$\\$\chi_\nu^2 = 0.3(1.4)$} \\
\bottomrule
\end{tabularx}

	\end{adjustwidth}
\end{table}
\begin{table}[H]
    \caption{As Table~\ref{tab:fss_chi_-0.1} but for $\J2=-0.21$.\label{tab:fss_chi_-0.21}}
	\begin{adjustwidth}{-\extralength}{0cm}
	\begin{tabularx}{\fulllength}{llllll}
\toprule
$\mathbf{L_{\max}}$ & 24 & 32 & 48 & 64 & 88 \\
$\mathbf{L_{\min}}$ &  &  &  &  &  \\
\midrule
8 & \makecell[l]{$a = 8.488(44)$\\$y_h = 1.8873(13)$\\$\chi_\nu^2 = 5.2(3.3)$} & \makecell[l]{$a = 8.537(37)$\\$y_h = 1.8860(11)$\\$\chi_\nu^2 = 10.1(4.6)$} & \makecell[l]{$a = 8.627(55)$\\$y_h = 1.8839(16)$\\$\chi_\nu^2 = 16.2(7.7)$} & \makecell[l]{$a = 8.637(61)$\\$y_h = 1.8837(18)$\\$\chi_\nu^2 = 12.8(6.4)$} & \makecell[l]{$a = 8.669(95)$\\$y_h = 1.8830(25)$\\$\chi_\nu^2 = 12.6(8.2)$} \\
16 & -- & \makecell[l]{$a = 8.795(97)$\\$y_h = 1.8812(23)$\\$\chi_\nu^2 = 2.6(2.8)$} & \makecell[l]{$a = 8.819(92)$\\$y_h = 1.8807(22)$\\$\chi_\nu^2 = 1.5(1.5)$} & \makecell[l]{$a = 8.81(10)$\\$y_h = 1.8810(24)$\\$\chi_\nu^2 = 1.3(1.3)$} & \makecell[l]{$a = 8.79(15)$\\$y_h = 1.8813(33)$\\$\chi_\nu^2 = 1.3(1.8)$} \\
24 & -- & -- & \makecell[l]{$a = 8.89(12)$\\$y_h = 1.8797(26)$\\$\chi_\nu^2 = 1.4(2.1)$} & \makecell[l]{$a = 8.84(14)$\\$y_h = 1.8804(30)$\\$\chi_\nu^2 = 1.5(2.0)$} & \makecell[l]{$a = 8.79(20)$\\$y_h = 1.8812(39)$\\$\chi_\nu^2 = 1.5(2.1)$} \\
32 & -- & -- & -- & \makecell[l]{$a = 8.68(17)$\\$y_h = 1.8828(33)$\\$\chi_\nu^2 = 0.23(83)$} & \makecell[l]{$a = 8.72(22)$\\$y_h = 1.8823(43)$\\$\chi_\nu^2 = 0.31(86)$} \\
48 & -- & -- & -- & -- & \makecell[l]{$a = 8.72(28)$\\$y_h = 1.8823(50)$\\$\chi_\nu^2 = 0.3(1.2)$} \\
\bottomrule
\end{tabularx}

	\end{adjustwidth}
\end{table}
\begin{table}[H]
    \caption{As Table~\ref{tab:fss_chi_-0.1} but for $\J2=-0.22$.\label{tab:fss_chi_-0.22}}
	\begin{adjustwidth}{-\extralength}{0cm}
	\begin{tabularx}{\fulllength}{llllll}
\toprule
$\mathbf{L_{\max}}$ & 24 & 32 & 48 & 64 & 88 \\
$\mathbf{L_{\min}}$ &  &  &  &  &  \\
\midrule
8 & \makecell[l]{$a = 11.203(30)$\\$y_h = 1.88778(65)$\\$\chi_\nu^2 = 5.2(3.7)$} & \makecell[l]{$a = 11.292(53)$\\$y_h = 1.8860(11)$\\$\chi_\nu^2 = 13.4(5.5)$} & \makecell[l]{$a = 11.373(67)$\\$y_h = 1.8845(12)$\\$\chi_\nu^2 = 21.6(8.4)$} & \makecell[l]{$a = 11.460(56)$\\$y_h = 1.88300(95)$\\$\chi_\nu^2 = 24.8(6.1)$} & \makecell[l]{$a = 11.467(55)$\\$y_h = 1.88288(95)$\\$\chi_\nu^2 = 20.6(5.0)$} \\
16 & -- & \makecell[l]{$a = 11.81(15)$\\$y_h = 1.8791(19)$\\$\chi_\nu^2 = 2.6(3.4)$} & \makecell[l]{$a = 11.92(13)$\\$y_h = 1.8777(16)$\\$\chi_\nu^2 = 2.2(2.4)$} & \makecell[l]{$a = 11.853(87)$\\$y_h = 1.8786(13)$\\$\chi_\nu^2 = 2.0(1.9)$} & \makecell[l]{$a = 11.851(83)$\\$y_h = 1.8786(12)$\\$\chi_\nu^2 = 1.5(1.4)$} \\
24 & -- & -- & \makecell[l]{$a = 12.07(12)$\\$y_h = 1.8758(16)$\\$\chi_\nu^2 = 0.15(75)$} & \makecell[l]{$a = 11.89(11)$\\$y_h = 1.8781(15)$\\$\chi_\nu^2 = 2.1(2.7)$} & \makecell[l]{$a = 11.89(10)$\\$y_h = 1.8782(15)$\\$\chi_\nu^2 = 1.5(1.9)$} \\
32 & -- & -- & -- & \makecell[l]{$a = 11.79(15)$\\$y_h = 1.8792(19)$\\$\chi_\nu^2 = 2.2(3.4)$} & \makecell[l]{$a = 11.79(14)$\\$y_h = 1.8792(18)$\\$\chi_\nu^2 = 1.2(1.7)$} \\
48 & -- & -- & -- & -- & \makecell[l]{$a = 11.54(27)$\\$y_h = 1.8817(30)$\\$\chi_\nu^2 = 0.8(1.9)$} \\
\bottomrule
\end{tabularx}

	\end{adjustwidth}
\end{table}
\begin{adjustwidth}{-\extralength}{0cm}

\reftitle{References}

\PublishersNote{}
\end{adjustwidth}